\documentclass[12pt]{article}

\usepackage[centertags]{amsmath}
\usepackage{amsfonts}
\usepackage{amssymb}
\usepackage{amsthm}
\usepackage{newlfont}
\usepackage{epsfig}
\usepackage{amscd}

\newcommand{\NN}{{\mathbb N}}

\newcommand{\RR}{{\mathbb R}}

\newcommand{\CC}{{\mathbb C}}
\newcommand{\beq}{\begin{equation}}
\newcommand{\eeq}{\end{equation}}
\newcommand{\ba}{\begin{array}}
\newcommand{\ea}{\end{array}}
\newcommand{\bea}{\begin{eqnarray}}
\newcommand{\eea}{\end{eqnarray}}

\begin{document}

\begin{center}
{\large \sc \bf On the solutions of the dKP equation:} 

\vskip 10pt 
{\large \sc \bf nonlinear Riemann Hilbert problem,} 

\vskip 10pt 

{\large \sc \bf longtime behaviour, implicit solutions} 

\vskip 10pt 
{\large \sc \bf and wave breaking}

\vskip 20pt

{\large  S. V. Manakov$^{1,\S}$ and P. M. Santini$^{2,\S}$}

\vskip 20pt

{\it 
$^1$ Landau Institute for Theoretical Physics, Moscow, Russia

\smallskip

$^2$ Dipartimento di Fisica, Universit\`a di Roma "La Sapienza", and \\
Istituto Nazionale di Fisica Nucleare, Sezione di Roma 1 \\
Piazz.le Aldo Moro 2, I-00185 Roma, Italy}

\bigskip

$^{\S}$e-mail:  {\tt manakov@itp.ac.ru, paolo.santini@roma1.infn.it}

\bigskip

{\today}

\end{center}

\begin{abstract}
We have recently solved the inverse scattering problem for one-parameter families   
of vector fields, and used this result to construct the formal solution of the Cauchy 
problem for a class of integrable nonlinear  
partial differential equations in multidimensions,  
including the second heavenly equation of Plebanski and the dispersionless Kadomtsev 
- Petviashvili (dKP) equation. We showed, in particular, that 
the associated inverse problems can be expressed 
in terms of nonlinear Riemann - Hilbert problems on the real axis.  
 In this paper we make use of the nonlinear Riemann - Hilbert problem of dKP,  i) to 
construct the longtime behaviour of the solutions of its Cauchy problem; 
ii) to characterize a class of implicit solutions;  
iii) to elucidate the spectral mechanism causing the gradient catastrophe
 of localized solutions of dKP, at finite time as well as in the longtime regime, and 
the corresponding universal behaviours near breaking. 
 
\end{abstract}

\newpage
\section{Introduction}
It was observed long ago \cite{ZS} that the commutation of multidimensional 
vector fields can generate integrable nonlinear partial differential equations (PDEs) in arbitrary 
dimensions. An important example of PDE of this type is the dispersionless Kadomtsev-Petviashvili 
(dKP) equation
\beq
\label{dKP}
(u_t+uu_x)_x+u_{yy}=0,~~~
u=u(x,y,t)\in\RR,~~~~~x,y,t\in\RR,
\eeq 
arising from the commutation
\beq
[\hat L_1,\hat L_2]=0
\eeq
of the following pair of $1$-parameter families of vector fields:
\beq
\label{L1L2}
\ba{l}
\hat L_1\equiv \partial_y+\lambda\partial_x-u_x\partial_{\lambda}, \\
\hat L_2\equiv \partial_t+(\lambda^2+u)\partial_x+(-\lambda u_x+u_y)\partial_{\lambda},
\ea
\eeq
$\lambda$ being the spectral parameter. 

The dKP equation is the $x$-dispersionless limit of the celebrated Kadomtsev-Petviashvili (KP)    
equation \cite{KP} $(u_t+uu_x+u_{xxx})_x+\sigma u_{yy}=0,~\sigma =\pm 1$, and therefore it describes, 
for instance,  
the evolution of small amplitude, nearly one-dimensional waves in shallow water \cite{AC} 
near the shore, when the $x$-dispersion can be neglected.   
It is also a model equation in the description of unsteady motion in transonic flow 
\cite{Timman} and in the nonlinear acoustics of confined beams \cite{KZ}.

We remark that, if also the $y$ - dispersion term $u_{yy}$ is negligeable in (\ref{dKP}), 
the dKP equation  reduces to the celebrated ($1+1$) - dimensional Hopf equation: 
$u_t+uu_x=0$, the integrable prototype model for 
the description of the gradient catastrophe (breaking) of a localized one - dimensional wave. 
Therefore a natural  
question is whether the dKP equation can be viewed as the integrable universal model in the 
description of the gradient catastrophe of a localized two - dimensional wave. As far as we know, 
no precise results are known in this direction, although some formal considerations have been made, 
in the framework of unsteady motions in transonic flow \cite{KC}. 
The following issues are, in particular, relevant. 
i) Do localized initial data evolving according to dKP 
break? ii) Do small initial data also break? ii) If breaking occurs, does it take place in a point 
of the $(x,y)$ - plane or on a line? iii) Do the analytic and geometric aspects  
of the breaking exhibit universal features, like in the one-dimensional case? iv) 
How are the breaking features connected with the dKP 
initial data? In this paper all these questions will find a proper answer.  

Dispersionless (or quasi-classical) limits 
of integrable PDEs, having dKP as prototype example, arise in various problems of 
Mathematical Physics and  are intensively studied in the recent literature 
(see, f.i., \cite{KG} - \cite{KM}). The Lax representation (\ref{L1L2}) and the 
Hamiltonian formulation for dKP can be found, for instance, in \cite{Zak} 
and in \cite{Kri}. An elegant integration scheme for dKP, applicable in general 
to nonlinear PDEs associated with Hamiltonian 
vector fields, was presented in \cite{Kri}. A nonlinear $\bar\partial$ - 
dressing was developed in \cite{K-MA-R}. Special classes of nontrivial solutions were also derived 
(see, f.i., \cite{DT}, \cite{G-M-MA}).  
 
The inverse scattering 
transform for $1$-parameter families of multidimensional vector fields has been 
developed in \cite{MS1} (see also \cite{MS1bis}). This theory, introducing interesting novelties 
with respect to the classical inverse scattering transform for soliton equations \cite{ZMNP,AC}, 
has allowed one to construct the formal solution of the Cauchy problems for the  
heavenly equation  \cite{Pleb} in \cite{MS1} and for the following novel system of PDEs
\beq
\label{dKP-system}
\ba{l}
u_{xt}+u_{yy}+(uu_x)_x+v_xu_{xy}-v_yu_{xx}=0,         \\
v_{xt}+v_{yy}+uv_{xx}+v_xv_{xy}-v_yv_{xx}=0,
\ea
\eeq
arising from the commutation of the vector fields 
\beq\label{Ltilde}
\ba{l}
\tilde L_1\equiv \partial_y+(\lambda+v_x)\partial_x-u_x\partial_{\lambda}, \\
\tilde L_2\equiv \partial_t+(\lambda^2+\lambda v_x+u-v_y)\partial_x+(-\lambda u_x+u_y)\partial_{\lambda}, 
\ea
\eeq
in \cite{MS3}. The Cauchy problem for the $v=0$ reduction of (\ref{dKP-system}), the dKP equation (\ref{dKP}), 
was also presented in \cite{MS3}, while the Cauchy problem for the 
$u=0$ reduction of (\ref{dKP-system}), an integrable system introduced in \cite{Pavlov2}, was given in \cite{MS4}.   
Due to the ring property 
of the space of eigenfunctions associated with vector fields, the inverse problem 
can be formulated in three different distinguished ways, one of them being expressed in terms of 
a linear integral equation. 

In this paper we show that one of these 
formulations, the nonlinear Riemann - Hilbert (RH) problem on the real axis,   
is an efficient way i) to construct 
the longtime behaviour of the solutions of the Cauchy problem for dKP; ii) to characterize 
a class of implicit solutions of dKP, parametrized by  an arbitrary spectral function of a single 
variable; iii) to establish that localized solutions of dKP generically break (if small, they break 
in the longtime regime), and to study in great detail the universal features of the wave breaking: 
the similarity solution before breaking, 
the vertical inflection at breaking, and the development of a three-valued region in the $(x,y)$ - plane 
delimited by a caustic after breaking; iv) to connect the analytic and universal features 
of this gradient catastrophe with the initial data, via the spectral transform  developed in \cite{MS3}.
 
The paper is organized as follows. In \S 2 we present the nonlinear RH dressing 
of the system (\ref{dKP-system}) and of its dKP reduction (\ref{dKP}), showing that it provides an implicit spectral 
representation of the solution 
depending on the solution itself, leading to the breaking of localized solutions at finite $t$. 
In \S 3 we make use of such a 
RH problem to construct the longtime behaviour of the solutions of the Cauchy 
problem for the dKP equation, showing that, also asymptotically, the above spectral mechanism 
for breaking is present.  
In \S 4 we select a class of spectral data for wich the vector RH problem decouples and 
linearizes, leading to a class of implicit solutions of dKP. In \S 5 we study the analytic 
aspects of the longtime breaking of a localized initial condition, 
showing that this problem is connected to the wave breaking of a localized 
two-dimensional wave evolving according to the 1+1 dimensional Hopf equation. In 
\S 6 we describe the analytic aspects of the wave breaking at finite time.

\section{Nonlinear RH inverse problem}

From the inverse scattering transform developed in \cite{MS3} for the system (\ref{dKP-system}) and for 
its dKP reduction (\ref{dKP}), one extracts the following inverse problem. 

\vskip 5pt
\noindent
{\it Proposition} Consider the vector nonlinear RH problem on the real line:  
\beq\label{RH}
\vec\pi^+(\lambda)=\vec {\cal R}(\vec\pi^-(\lambda)),~~\;\; \lambda\in\RR,
\eeq
where the solutions $\vec\pi^{\pm}(\lambda)=(\pi^{\pm}_1(\lambda),\pi^{\pm}_2(\lambda))\in\CC^2$ 
are $2$-dimensional vector functions analytic in the upper and lower 
halves of the complex $\lambda$ plane, normalized in the following way
\beq
\label{pi_asympt_1}
\vec\pi^{\pm}(\lambda)=\left(
\ba{c}
-\lambda^2t-\lambda y+x-2ut \\
\lambda
\ea
\right)+\vec O(\lambda^{-1}),~~~|\lambda |>>1, 
\eeq
where
\beq\label{def_u}
u=\lim\limits_{\lambda\to\infty}{\big(\lambda(\pi^{\pm}_2(\lambda)-\lambda})\big),
\eeq 
and $\vec {\cal R}(\vec \zeta)=({\cal R}_1(\zeta_1,\zeta_2),{\cal R}_2(\zeta_1,\zeta_2))\in\CC^2,~
\vec \zeta\in\CC^2$ are given differentiable spectral data. 

Then, assuming that the above RH problem and its linearized form $\vec\sigma^+=J\vec\sigma^-$ are 
uniquely solvable,  
where $J$ is the Jacobian matrix of the transformation (\ref{RH}): 
$J_{ij}=\partial {\cal R}_i/\partial\zeta_j,~i,j=1,2$, the solutions $\vec\pi^{\pm}(\lambda)$ of the 
RH problem (\ref{RH}) are common eigenfunctions of the vector fields (\ref{Ltilde}): 
$\tilde {L}_j\vec\pi^{\pm}=\vec 0,~~~~j=1,2$, where 
\beq\label{def_v}
v(x,y,t)=-yu-\lim\limits_{\lambda\to\infty}{\lambda\left(\pi^{\pm}_1+t{\pi^{\pm}_2}^2-(x-\lambda y)\right)},
\eeq
and $u,v$ are solutions of the nonlinear system (\ref{dKP-system}).

If the spectral data $\vec {\cal R}(\vec \zeta)$ satisfy the reality constraint
\beq\label{R_reality}
\ba{l}
\vec{\cal R}(\overline{ \vec{\cal R}(\bar{\vec\zeta})})=\vec\zeta,
~~~\forall\vec\zeta\in\CC^2,
\ea
\eeq
then the solutions $u,v$ are real: $u,v\in \RR$. If, in addition, the transformation 
$\vec \zeta \to \vec {\cal R}(\vec \zeta)$ is canonical:
\beq\label{R_dKP}
\ba{l}
\{{\cal R}_1,{\cal R}_2\}_{\vec\zeta}=1,
\ea
\eeq
where $\{\cdot,\cdot\}$ is the Poisson bracket with respect to the variables $(\zeta_1,\zeta_2)$,  
then $v=0$, $\vec\pi^{\pm}(\lambda)$  are common eigenfunctions 
of the vector fields (\ref{L1L2}): $\hat{L}_j\vec\pi^{\pm}=\vec 0,~j=1,2$  and $u$ 
satisfies the dKP equation (\ref{dKP}).

\vskip 5pt
\noindent
{\it Proof}. One easily shows that the vectors $\tilde L_j\vec\pi^{\pm},~j=1,2$ solve the linearized RH problem: 
$(\tilde L_j\vec\pi^{+})=J(\tilde L_j\vec\pi^{-})$, and that $\tilde L_j\vec\pi^{\pm}\to 0$ as $\lambda\to\infty$.   
By uniqueness, we infer that $\vec\pi^{\pm}$ are eigenfunctions 
of the operators $\tilde L_j$: $\tilde L_j\vec\pi^{\pm}=0,~j=1,2$, implying that $u$ and $v$ are 
solutions of the nonlinear system (\ref{dKP-system}).

From the reality constraint (\ref{R_reality}) it follows that 
$\overline{\vec\pi^{+}}=\vec\pi^{-}$, implying $u,v\in\RR$. From the dKP constraint (\ref{R_dKP}) 
it follows that $\{\pi^{+}_1,\pi^{+}_2\}=\{\pi^{-}_1,\pi^{-}_2\},~\lambda\in\RR$, where $\{\cdot,\cdot\}$ 
is now the Poisson bracket with respect to the variables $(x,\lambda)$. Since 
$\{\pi^{\pm}_1,\pi^{\pm}_2\}\to 1$, as $\lambda\to\infty$, the analyticity properties 
of the eigenfunctions imply that 
$\{\pi^{+}_1,\pi^{+}_2\}=\{\pi^{-}_1,\pi^{-}_2\}=1$. At last, applying $\tilde L_{j},~j=1,2$ to this 
equation, on obtains the Hamiltonian constraint $v=0$, and system (\ref{dKP-system}) reduces 
to the dKP equation. $\Box$ 

\vskip 10pt
From the integral equation 
\beq\label{pi_equ}
\ba{l}
\pi^-_1(\lambda)=-\lambda^2t-\lambda y+x-2ut+
\frac{1}{2\pi i}\int\limits_{\RR}\frac{d\lambda'}{\lambda'-(\lambda -i0)}
R_1\Big(\pi^-_1(\lambda'),\pi^-_2(\lambda')\Big), \\
\pi^-_2(\lambda)=\lambda+\frac{1}{2\pi i}\int\limits_{\RR}\frac{d\lambda'}{\lambda'-(\lambda -i0)}
R_2\Big(\pi^-_1(\lambda'),\pi^-_2(\lambda')\Big),
\ea
\eeq 
characterizing the solutions of the RH problem (\ref{RH}), and from the definition (\ref{def_u}),   
one obtains the following spectral characterization of the solution $u$:
\beq\label{inverse_1}
u=F\left(x-2ut,y,t\right)\in\RR,
\eeq
where the spectral function $F$, defined by 
\beq\label{inverse_2}
\ba{l}
F\left(\xi,y,t \right)=-\int\limits_{\RR}\frac{d\lambda}{2\pi i}R_2
\Big(\pi^-_1(\lambda;\xi,y,t),\pi^-_2(\lambda;\xi,y,t)\Big),  \\
{\cal R}_j(\zeta_1,\zeta_2)=\zeta_j+R_j(\zeta_1,\zeta_2),~~~j=1,2,
\ea
\eeq
is connected with the initial data $u(x,y,0)$ via the direct spectral transform 
developed in \cite{MS3}.

We remark that, while the inverse spectral transform of most of the known integrable PDEs provides 
a spectral representation of the solution involving, as parameters, the space-time 
coordinates, the inverse problem of dKP provides a spectral representation 
(\ref{inverse_1}), (\ref{inverse_2}) of the solution involving, as parameter,   
also the solution $u$ itself, in the combination $(x-2ut)$. This is the spectral mechanism 
for the breaking of a generic localized initial condition at finite time $t$. We postpone to \S 5,6 a 
detailed analysis of the analyticity aspects of such a breaking. 

We also remark that the mechanism responsible for this feature is that the vector field $\hat L_2$ 
is quadratic in $\lambda$ and, at the same time, it contains the partial differential operator 
$\partial_{\lambda}$. Due to that, the normalization of the analytic eigenfunction of $\hat L_2$ 
involves the coefficient $u$ of the vector field. It is easy to see that these properties 
are shared by the whole dKP hierarchy, associated with time operators 
$\hat L_n$ involving higher powers of $\lambda$. An example of 
integrable system not associated with the commutation of vector fields, but exhibiting 
an inverse problem in which the solution is implicit, is the Harry 
Dym equation (\cite{KL,DK}), connected with the KdV equation by a hodograph transformation.

\section{Asymptotics of the dKP equation}

In this section we 
concentrate on the study of the longtime behaviour of the solutions of dKP. 
Of course, this study is meaningful only if no breaking  
takes place before, at finite time. If the initial condition is, for instance, small, 
the nonlinear term in (\ref{dKP}) becomes important only in the longtime 
regime, and no breaking takes place before.    

As we shall see in the following, the spectral mechanism causing the  
breaking of a localized initial condition evolving according to the dKP equation 
is present also in the longtime regime.    

We study the longtime $t>>1$ regime in the space regions
\beq\label{asympt_reg}
x=\tilde x+v_1t,~~y=v_2t,~~~~~\tilde x -2ut,v_1,v_2=O(1),~~v_2\ne 0,~~t>>1.
\eeq  

Then the system of nonlinear integral equations (\ref{pi_equ}) 
is conveniently rewritten in terms of the functions $\phi_j,~j=1,2$ 
defined by
\beq\label{def_phi}
\phi_1(\lambda)=\pi^-_1(\lambda)+\lambda^2t+\lambda y-(x-2ut),~~~
\phi_2(\lambda)=\pi^-_2(\lambda)-\lambda,
\eeq
as follows
\beq\label{phi_equ}
\ba{l}
\phi_j(\lambda)= 
\frac{1}{2\pi i}\int_{\RR}\frac{d\lambda'}{\lambda'-(\lambda -i0)}\times \\
R_j\Big(\tilde x-2ut-(\lambda'-\lambda_+)(\lambda'-\lambda_-)t+\phi_1(\lambda'),
\lambda'+\phi_2(\lambda')\Big),~j=1,2, 
\ea
\eeq 
where
\beq
\lambda_{\pm}=-\frac{v_2}{2}\pm\sqrt{v_1+\frac{v^2_2}{4}}.
\eeq
The fast decay, in $t$, of $\phi_j,~j=1,2$, due to the  
linear growth in $t$ of the first argument of $R_j,~j=1,2$, is partially contrasted 
if $\lambda_+=\lambda_-$; i.e., on the parabola:
\beq\label{parabola}
x+\frac{y^2}{4t}=\tilde x~~~~~~\left(v_1+\frac{v^2_2}{4}=0\right)
\eeq
of the $(x,y)$ - plane. 

On such parabola the integral equations read
\beq\label{phi2_equ_heav}
\ba{l}
\phi_j(\lambda)= 
\frac{1}{2\pi i}\int_{\RR}\frac{d\lambda'}{\lambda'-(\lambda -i0)}\times   \\
R_j\Big(\tilde x-2ut-\left(\lambda'+\frac{v_2}{2}\right)^2t+\phi_1(\lambda'),
\lambda'+\phi_2(\lambda')\Big),~j=1,2.
\ea
\eeq 
Since, in this case, the main contribution to the integrals occurs when $\lambda'\sim -v_2/2$, 
we make the change of variable $\lambda'=-v_2/2+\mu'/\sqrt{t}$, obtaining:
\beq\label{phi3_equ_heav}
\ba{l}
\phi_j(\lambda)= 
\frac{1}{2\pi i\sqrt{t}}\int\limits_{\RR}\frac{d\mu'}{\mu'/\sqrt{t}-(\lambda+v_2/2-i0)}
R_j\Big(\tilde x-2ut-{\mu'}^2+\phi_1(-\frac{v_2}{2}+\frac{\mu'}{\sqrt{t}}), \\
-\frac{v_2}{2}+\frac{\mu'}{\sqrt{t}}+\phi_2(-\frac{v_2}{2}+\frac{\mu'}{\sqrt{t}})\Big),~j=1,2.
\ea
\eeq
If $|\lambda+v_2/2|>>t^{-1/2}$, equation (\ref{phi3_equ_heav}) implies that 
$\phi_{j}(\lambda)=O(t^{-1/2}),~j=1,2$: 
\beq\label{phi_asympt1_heav}
\ba{l}
\phi_j(\lambda)\sim 
-\frac{1}{2\pi i\sqrt{t}(\lambda +v_2/2-i0))}\int\limits_{\RR}d\mu'
R_j\Big(\tilde x-2ut-{\mu'}^2+\phi_1(-\frac{v_2}{2}+\frac{\mu'}{\sqrt{t}}), \\
-\frac{v_2}{2}+\phi_2(-\frac{v_2}{2}+\frac{\mu'}{\sqrt{t}}) \Big),~j=1,2,
\ea
\eeq
and, via (\ref{inverse_1})-(\ref{inverse_2}), that
\beq
\ba{l}
u=-\frac{1}{2\pi i\sqrt{t}}
\int\limits_{\RR}d\mu R_2\Big(\tilde x-2ut-{\mu'}^2+\phi_1(-\frac{v_2}{2}+\frac{\mu'}{\sqrt{t}}), \\
-\frac{v_2}{2}+\phi_2(-\frac{v_2}{2}+\frac{\mu'}{\sqrt{t}}) \Big)+o\left(\frac{1}{\sqrt{t}}\right).
\ea
\eeq
For $\lambda +v_2/2=\mu /\sqrt{t},~\mu=O(1)$, instead, $\phi_{j}(\lambda),~j=1,2$ are $O(1)$:
\beq\label{phi_asympt2_heav}
\ba{l}
\phi_j(-\frac{v_2}{2}+\frac{\mu}{\sqrt{t}})\sim 
\frac{1}{2\pi i}\int\limits_{\RR}\frac{d\mu'}{\mu'-(\mu-i0)}
R_j\Big(\tilde x-2ut-{\mu'}^2+\phi_1(-\frac{v_2}{2}+\frac{\mu'}{\sqrt{t}}), \\
-\frac{v_2}{2}+\phi_2(-\frac{v_2}{2}+\frac{\mu'}{\sqrt{t}} \Big),~~~j=1,2. 
\ea
\eeq
Therefore it is not possible to neglect, in the integral equations (\ref{phi3_equ_heav}), 
$\phi_{1,2}$ in the arguments of $R_{1,2}$; it follows that these integral equations remain nonlinear also 
in the longtime regime.  

The above results can be summarized as follows. 

\noindent
On the parabola (\ref{parabola}), in the space-time regions (\ref{asympt_reg}), 
the longtime behaviour of the solution of the dKP equation is given by
\beq
\label{asympt_dKP_1}
u= \frac{1}{\sqrt{t}}G\left(x+\frac{y^2}{4t}-2ut,\frac{y}{2t}\right)+
o\left(\frac{1}{\sqrt{t}}\right), 
\eeq
where
\beq
\label{asympt_dKP_2}
G(\xi,\eta)=
-\frac{1}{2\pi i}\int\limits_{\RR}d\mu
R_2\Big(\xi-\mu^2+a_1(\mu;\xi,\eta),-\eta+a_2(\mu;\xi,\eta) \Big),
\eeq
$R_2$ is the second component of the spectral data $\vec R$, and $a_{1,2}(\mu;\xi,\eta)$  
are the unique solutions of the nonlinear integral equations
\beq\label{a_dKP}
\ba{l}
a_j(\mu;\xi,\eta)= 
\frac{1}{2\pi i}\int\limits_{\RR}\frac{d\mu'}{\mu'-(\mu-i0)}
R_j\Big(\xi-{\mu'}^2+a_1(\mu';\xi,\eta),   \\
-\eta+a_2(\mu';\xi,\eta) \Big),~~~~~j=1,2.
\ea
\eeq

We remark that the system of integral equations (\ref{a_dKP})   
characterizes the solutions of the following vector nonlinear RH problem 
on the real $\mu$ - axis: 
\beq\label{RH_asympt_dKP}
\ba{l}
\vec A^+(\mu;\xi,\eta)=\vec A^-(\mu;\xi,\eta)+\vec R(\vec A^-(\mu;\xi,\eta)),~~
\mu\in\RR, \\
\vec A^{\pm}(\mu;\xi,\eta)=\left(
\ba{c}
\xi- \mu^2 \\
-\eta
\ea
\right) +\vec O(\mu^{-1}),~~|\mu|>>1,  
\ea
\eeq
with
\beq
\vec a(\mu;\xi,\eta)=\vec A^-(\mu;\xi,\eta)-
\left(
\ba{c}
\xi- \mu^2 \\
-\eta
\ea
\right) .
\eeq
Outside the parabola, the solution decays faster than $1/\sqrt{t}$.

We remark that equation (\ref{asympt_dKP_1}) has been obtained under the hypothesis that both arguments of 
$G$ are $O(1)$; then $u=O(1/\sqrt{t})$. It follows that $2ut=O(\sqrt{t})$ and, 
consequently, also $\tilde x=x+y^2/(4t)=O(\sqrt{t})$, to balance the term $2ut$.

From equation (\ref{asympt_dKP_1}) it follows that the spectral mechanism causing the  
breaking of a localized initial condition evolving according to the dKP equation 
is present also in the longtime regime.  The analytic aspects of the longtime breaking 
of dKP solutions are illustrated in \S 5.

\section{Particular solutions of the RH problem}

In this section we construct a class of explicit solutions of the vector nonlinear RH problem 
(\ref{RH}) and, correspondingly, a class of implicit solutions
 of the dKP equation parametrized by an arbitrary real spectral function of one variable (the 
general solution would depend on an arbitrary real spectral function of two variables). 

Suppose that the components of the spectral data $\vec{\cal R}$ in (\ref{RH}) are given by:
\beq
{\cal R}_j(\zeta_1,\zeta_2)=\zeta_j e^{i(-)^{j+1}f(\zeta_1\zeta_2)}-1,~~j=1,2,
\eeq
in terms of the single real spectral function $f$ of a single argument, depending on 
$\zeta_1$ and $\zeta_2$ only through their product. 

Then the RH problem becomes   
\beq\label{RH_heav_part}
\ba{l}
\pi^+_1=\pi^-_1e^{if\left(\pi^-_1\pi^-_2\right)},~~~\lambda\in\RR, \\
\pi^+_2=\pi^-_2e^{-if\left(\pi^-_1\pi^-_2\right)},
\ea 
\eeq
and the following properties hold.

\vskip 10pt
\noindent
i) The reality and the dKP constraints (\ref{R_reality}),(\ref{R_dKP}) are satisfied.

\vskip 10pt
\noindent
ii) $\pi^+_1\pi^+_2=\pi^-_1\pi^-_2$. Consequently, ($\pi^+_1\pi^+_2$) is just a 
polynomial in $\lambda$:
\beq\label{w_dKP}
\pi^+_1\pi^+_2=\pi^-_1\pi^-_2=-t\lambda^3-y\lambda^2+(x-3ut)\lambda -2y u+3t\;\partial^{-1}_xu_y
 \equiv w(\lambda),
\eeq
and the vector nonlinear RH problem (\ref{RH}) decouples into two  
scalar, linear RH problems:
\beq\label{RH_linear}
\ba{l}
\pi^+_1=\pi^-_1e^{if(w(\lambda))}, \\
\pi^+_2=\pi^-_2e^{-if(w(\lambda))}.
\ea
\eeq

\vskip 10pt
\noindent
iii) Since, from (\ref{RH_linear}), 
\beq
\pi^+_je^{i(-)^{j}f^+(\lambda)}=\pi^-_je^{i(-)^{j}f^-(\lambda)},~~j=1,2,
\eeq
where
\beq
f^{\pm}(\lambda)=\frac{1}{2\pi i}
\int_{\RR}\frac{d\lambda'}{\lambda'-(\lambda\pm i0)}f(w(\lambda')),
\eeq
also $\pi^+_je^{i(-)^{j}f^+(\lambda)},~j=1,2$ are polynomials in $\lambda$.  
We expand them in powers of $\lambda$ and introduce the notation 
\beq
\label{momenta}
<\lambda^n f>=\frac{1}{2\pi}\int_{\RR}\lambda^n f(w(\lambda))d\lambda,~~n\in\NN .
\eeq 
From the positive power expansion it follows that
\beq
\label{bb}
\ba{l}
\pi^+_1e^{-if^+(w(\lambda))}=\pi^-_1e^{-if^-(w(\lambda))}=-t\lambda^2-(y+t<f>)\lambda + \\
x-2ut-y<f>-t\left( <\lambda f>+\frac{<f>^2}{2} \right) , \\
~~    \\
\pi^+_2e^{if^+(w(\lambda))}=\pi^-_2e^{if^-(w(\lambda))}=\lambda -<f>,
\ea
\eeq
implying the following explicit solution of the Riemann problem:
\beq\label{sol_dKP}
\ba{l}
\pi^{\pm}_1=
\Big[-\lambda^2t-\lambda (y+t<f>)+   \\
x-2ut-y<f>-t\left(<\lambda f>+\frac{<f>^2}{2}\right)\Big]\;e^{if^{\pm}(w(\lambda))}, \\
~~  \\
\pi^{\pm}_2=(\lambda-<f>) \;e^{-if^{\pm}(w(\lambda))}.
\ea
\eeq
From the $1/\lambda$ terms of $\pi^+_je^{i(-)^{j}f^+(\lambda)},~j=1,2$, that must be zero, 
one finally obtains the two conditions
\beq
\label{implicit_dKP}
\ba{l}
u=<\lambda f>-\frac{<f>^2}{2},  \\
~~  \\
2t\;\partial^{-1}_xu_y=yu-(x-2ut)<f>+y\left(<\lambda f>+\frac{<f>^2}{2}\right)+  \\
t\left(<\lambda^2 f>+<f><\lambda f>+\frac{<f>^3}{6}\right)
\ea
\eeq  

Since, through (\ref{w_dKP}) and (\ref{momenta}), $<\lambda^n f>,~n=0,1,2$ depend 
implicitely on the solution $u$ and on $(t\;\partial^{-1}_xu_y)$, the system (\ref{implicit_dKP}) 
characterizes a class of implicit solutions 
of the dKP equation, parametrized by the arbitrary real spectral function $f$ of a single variable. 
    
\section{Longtime breaking of dKP solutions}

In this section we show that longtime solutions of dKP break. Of course this 
analysis is meaningful only when breaking does not take place before, at finite time. If,  
for instance, the localized initial condition is small, the nonlinear 
term of the dKP equation becomes relevant only in the longtime regime, and breaking can occur only 
when $t$ is large. 

Let $U(x,y,t)$ be the exact solution of the functional equation (\ref{asympt_dKP_1}) 
at the leading order; i.e.:
\beq\label{U}
U(x,y,t)=\frac{1}{\sqrt{t}}G\left(x+\frac{y^2}{4t}-2\;U\;t,\frac{y}{2t}\right),
\eeq
where $G$, connected with the initial condition of dKP via the direct spectral problem 
introduced in \cite{MS3}, is a largely arbitrary differentiable function of two arguments. 
It is easy to verify that 
$U$ is the general solution of the first order PDE in $2+1$ dimensions:
\beq
\label{U_equ}
U_t+\frac{y}{t}U_y-\frac{y^2}{4t^2}U_x+\frac{U}{2t}+UU_x=0.
\eeq
Equation (\ref{U}) suggests to introduce the convenient variables:
\beq\label{change_variables}
\ba{l}
V=\sqrt{t}\;U, \\
\tilde x= x+\frac{y^2}{4t},~~\tilde y=\frac{y}{2t},~~\tilde t=2\sqrt{t},
\ea
\eeq
transforming the PDE (\ref{U_equ}) into the $1+1$ dimensional Hopf equation:
\beq\label{Hopf}
V_{\tilde t}+VV_{\tilde x}=0.
\eeq
Then {\bf the longtime behaviour of the dKP solutions is reduced to the study of the evolution 
of a two-dimensional localized wave under the $1+1$ dimensional Hopf equation 
(\ref{Hopf})}; its solution, defined implicitly by the equations
\beq\label{sol_Hopf}
\ba{l}
V=G(\xi, \tilde y), \\
\xi=\tilde x - G(\xi,\tilde y)\tilde t,
\ea
\eeq
depends on $\tilde y$ only parametrically. 

Each small portion of the two-dimensional localized 
wave, characterized by its own amplitude $V$, travels with constant speed $V$  
in the $\tilde x$ - direction. Equations (\ref{sol_Hopf}b) define a 2-parameter family 
of characteristic curves in the $(\tilde x,\tilde y,\tilde t)$-space, the parameters being $\xi,\tilde y$. 
Due to the localisation properties of $G$, on 
each plane $\tilde y =$ const, the characteristic curves obtained varying $\xi$ intersect,  
giving rise to the breaking phenomenon. The first breaking will take place therefore 
in a specific point $(\tilde x_b,\tilde y_b,\tilde t_b)$ and, going back to the physical  
variables inverting the transformation (\ref{change_variables}), in a specific point $(x_b,y_b)$ of the 
$(x,y)$ - plane, at the breaking time $t_b$ (see Fig.1).   

\subsection{2D - breaking according to the Hopf equation}

In this section we discuss in detail the analytic aspects of the wave breaking 
of a localized two-dimensional wave evolving according to the Hopf equation (\ref{Hopf}), 
whose solution is defined implicitely by equations (\ref{sol_Hopf}). 

One solves (\ref{sol_Hopf}b) 
with respect to the parameter $\xi$, obtaining $\xi(\tilde x,\tilde y,\tilde t),$ 
and replaces it into (\ref{sol_Hopf}a), to obtain the solution 
$V=G(\xi(\tilde x,\tilde y,\tilde t),\tilde y)$. The inversion of equation (\ref{sol_Hopf}b) 
is possible iff its $\xi$-derivative is different from zero. Therefore the singularity manifold (SM) of 
the two dimensional Cauchy problem for the Hopf equation is the two - dimensional manifold 
characterized by the equation 
\beq\label{SM_Hopf}
{\cal S}(\xi,\tilde y,t)\equiv 1+G_{\xi}(\xi,\tilde y)\tilde t=0~~~
\Rightarrow~~~\tilde t=-\frac{1}{G_{\xi}(\xi,\tilde y)}.
\eeq
Since 
\beq
\label{gradV}
\ba{l}
\nabla_{(\tilde x,\tilde y)}V=\frac{\nabla_{(\xi,\tilde y)}G(\xi,\tilde y)}{1+G_{\xi}(\xi,\tilde y)\tilde t},
\ea
\eeq
the slopes of the localized wave become infinity (the so-called gradient catastrophe) on the SM, 
and the two-dimensional wave ``breaks''. 
 
Then the first breaking time $\tilde t_b$, and the corresponding characteristic parameters  
${\vec\xi}_b=(\xi_b,\tilde y_b)$ are defined by 
\beq
\label{tilde_t_b}
\tilde t_b=-\frac{1}{G_{\tilde \xi}({\vec \xi}_b)}=
\mbox{global min}\left(-\frac{1}{G_{\xi}(\xi,\tilde y)}\right)>0,
\eeq 
and characterized by the equations:
\beq\label{def_breaking}
\ba{l}
G_{\xi}({\vec \xi}_b)<0,~~~~~~~~G_{\xi\xi}({\vec \xi}_b)=G_{\xi\tilde y}({\vec \xi}_b)=0, \\
G_{\xi\xi\xi}({\vec \xi}_b)>0,~~~~~~
\alpha \equiv G_{\xi\xi\xi}({\vec \xi}_b)G_{\xi\tilde y\tilde y}({\vec \xi}_b)- 
G^2_{\xi\xi\tilde y}({\vec \xi}_b)>0.
\ea
\eeq

The corresponding point $\tilde{\vec x}_b=(\tilde x_b,\tilde y_b)$ in which the first wave breaking 
takes place is, from (\ref{sol_Hopf}b):
\beq\label{tilde_x_b}
\tilde x_b= \xi_b + G({\vec \xi}_b)\tilde t_b.
\eeq 

Now we evaluate equations (\ref{sol_Hopf}b) and (\ref{SM_Hopf}) near breaking, in the regime:
\beq
\label{near_breaking_Hopf}
\tilde x=\tilde x_b+\tilde x',~~\tilde y=\tilde y_b+\tilde y',~~\tilde t=\tilde t_b+\tilde t',~~\xi=\xi_b+{\xi'},
\eeq
where $\tilde x',\tilde y',\tilde t',{\xi'}$ are small. Using (\ref{tilde_t_b}) - (\ref{tilde_x_b}), 
equation (\ref{sol_Hopf}b) becomes, at the leading order, 
the following cubic equation in ${\xi'}$:
\beq\label{cubic}
{\xi'}^3+a(\tilde y'){\xi'}^2+b(\tilde y',\tilde t'){\xi'}-\gamma X(\tilde x',\tilde y',\tilde t')=0,
\eeq
where
\beq\label{def_a,b,X}
\ba{l}
a(\tilde y')=\frac{3G_{\xi\xi\tilde y}}{G_{\xi\xi\xi}}\tilde y',~~~~
b(\tilde y',\tilde t')=
\frac{3}{G_{\xi\xi\xi}}\left[G_{\xi}\epsilon +G_{\xi\tilde y\tilde y}\tilde {y'}^2 \right], \\
X(\tilde x',\tilde y',\tilde t')=\tilde x'-G(\xi_b,\tilde y_b+\tilde y')\tilde t'-
\left[G(\xi_b,\tilde y_b+\tilde y')-G\right]\tilde t_b  \sim \\
\tilde x'+\frac{G_{\tilde y}}{G_{\xi}}\tilde y'-G \tilde t'+\frac{G_{\tilde y\tilde y}}{2G_{\xi}}\tilde {y'}^2-
G_{\tilde y}{\tilde y'}\tilde t'+\frac{G_{\tilde y\tilde y\tilde y}}{6G_{\xi}}\tilde {y'}^3,~~~~
\gamma=\frac{6|G_{\xi}|}{G_{\xi\xi\xi}},
\ea
\eeq
where 
\beq\label{epsilon}
\epsilon\equiv \frac{2\tilde t'}{\tilde t_b},
\eeq
corresponding to the maximal balance
\beq\label{max_balance_Hopf}
|{\xi'}|,|\tilde y'|=O(|\epsilon|^{1/2}),~~~|X|=O(|\epsilon|^{3/2}).
\eeq
In (\ref{def_a,b,X}) and in the rest of this section, all partial derivatives of $G$ whose 
arguments are not indicated are meant to be evaluated at $\vec\xi_b=(\xi_b,\tilde y_b)$.

The three roots of the cubic are given by the well-known Cardano's formula:
\beq\label{sol_cubic1}
\ba{l}
{\xi'}_0\left(\tilde x',\tilde y',\tilde t'\right)=-\frac{a}{3}+(A_+)^\frac{1}{3} + (A_-)^\frac{1}{3} , \\
{\xi'}_{\pm}\left(x',y',t'\right)=-\frac{a}{3}-\frac{1}{2}\left((A_+)^\frac{1}{3} + 
(A_-)^\frac{1}{3}\right) \pm \frac{\sqrt{3}}{2}i\left((A_+)^\frac{1}{3} - (A_-)^\frac{1}{3}\right),  
\ea
\eeq
where
\beq\label{def_A}
\ba{l}
A_{\pm}=R\pm \sqrt{\Delta}
\ea
\eeq
and the discriminant $\Delta$ reads
\beq
\label{discriminant}
\Delta=R^2+Q^3,
\eeq
with
\beq
\label{def_QR_Hopf} 
\ba{l}
Q(\tilde y',\tilde t')=\frac{3b-a^2}{9}=-\frac{|G_{\xi}|}{G_{\xi\xi\xi}}\epsilon + 
\frac{\alpha}{G^2_{\xi\xi\xi}}\tilde {y'}^2,    \\
R(\tilde x',\tilde y',\tilde t')=\frac{\gamma}{2}X(\tilde x',\tilde y',\tilde t')+\frac{ab}{18}+
\frac{a}{3}Q(\tilde y',\tilde t').
\ea
\eeq
At the same order, function $\cal S$ in (\ref{SM_Hopf}) becomes
\beq
\label{SM_Hopf_2}
\ba{l}
{\cal S}(\xi,\tilde y,t)=G_{\xi}\tilde t'+
\frac{1}{2}[G_{\xi\xi\xi}{\xi'}^2+2G_{\xi\xi\tilde y}{\xi'}\tilde y'+
G_{\xi\tilde y\tilde y}\tilde {y'}^2]\tilde t_b.
\ea
\eeq

Known ${\xi'}$ as function of ($\tilde x,\tilde y,\tilde t$) from the cubic (\ref{cubic}), 
the solution $V$ of the Hopf equation and its gradient are then approximated, near breaking, 
by the formulae:
\beq
\label{V_x}
\ba{l}
V(\tilde x,\tilde y,\tilde t)\sim G(\xi_b+{\xi'},\tilde y_b+\tilde y'),    \\
\nabla_{(\tilde x,\tilde y)}V\sim \frac{\nabla_{(\xi',\tilde y')}G(\xi_b+{\xi'},\tilde y_b+\tilde y')}
{G_{\xi}\tilde t'+\frac{1}{2}[G_{\xi\xi\xi}{\xi'}^2+2G_{\xi\xi\tilde y}{\xi'}\tilde y'+
G_{\xi\tilde y\tilde y}\tilde{y'}^2]\tilde t_b}.
\ea
\eeq
Together with the breaking point $\tilde{\vec x}_b$, another distinguished point is the inflection point 
$\tilde{\vec x}_f$:
\beq\label{tilde_inflection}
\tilde{\vec x}_f=(\tilde{x}_f(\tilde t'),\tilde y_b),~~~~\tilde{x}_f(\tilde t')=\tilde x_b+G\tilde t'
\eeq
at which $R=X=\tilde y'=a=\xi'=\xi'_{\tilde x\tilde x} \equiv \xi'_{\tilde x\tilde y}=0$ 
and, consequently:
\beq\label{flex}
\ba{l}
V=G,~~~ \nabla_{(\tilde x,\tilde y)}V=\frac{1}{\tilde t'}\left(1,\frac{G_{\tilde y}}{G_{\xi}}\right),~~~
V_{\tilde x\tilde x}=V_{\tilde x\tilde y}=0.
\ea
\eeq
We observe that $\tilde{\vec x}_f=\tilde{\vec x}_b$ at $\tilde t=\tilde t_b$.

\subsubsection{Before breaking} 

If $\tilde t<\tilde t_b$ $(\tilde t'<0)$, the coefficient $Q$, defined in (\ref{def_QR_Hopf}), 
is strictly positive, 
due to (\ref{def_breaking}); then the discriminant $\Delta=R^2+Q^3$ is also strictly positive  
and only the root $\xi'_0$ is real. Correspondingly, the real solution of (\ref{Hopf}) is 
single valued and described by Cardano's formula. In addition, function $\cal S$ in (\ref{SM_Hopf_2}) 
is also strictly positive and $\nabla_{(\tilde x,\tilde y)}V$ is finite $\forall ~\tilde x,\tilde y$. 

To have a more explicit solution, we restrict the asymptotic region around the inflection point 
to the narrow strip :
\beq
\label{strip_Hopf1}
\ba{l}
|\tilde y'|=O(\epsilon^{q}),~~|X(\tilde x',\tilde y',\tilde t')|=O(\epsilon^{p+1}), 
\ea
\eeq
where
\beq
\label{strip_Hopf2}
\ba{c}
\mbox{max}\left(\frac{p+1}{2},p\right)<q<p+1,~~~p>\frac{1}{2},  
\ea
\eeq
obtaining 
\beq
\label{aa}
\ba{l}
X\sim \tilde x'+(G_{\tilde y}/G_{\xi})\tilde y'-G\tilde t'=
\tilde {\vec k}\cdot (\tilde{\vec x}-\tilde{\vec x}_f(\tilde t'))=O(\epsilon^{p+1}),      \\
\xi' \sim \frac{\tilde x'+(G_{\tilde y}/G_{\xi})\tilde y'-G\tilde t'}{G_{\xi}\tilde t'}=O(\epsilon^{p}),
\ea
\eeq
where
\beq
\label{k}
\ba{l}
\tilde {\vec k}=\left(1,\frac{G_{\tilde y}}{G_{\xi}}\right)
\ea
\eeq
is normal to the strip and defines the breaking direction.
 
Replacing (\ref{aa}) into (\ref{V_x}) we obtain that, in the narrow strip (\ref{strip_Hopf1}) -
(\ref{strip_Hopf2}), the solution exhibits a universal behaviour, coinciding with the following 
 exact similarity solution of the Hopf equation (\ref{Hopf})  
\beq
\label{simil_asymp_1}
\ba{l}
V\sim 
\frac{\tilde x-\tilde x_b +(G_{\tilde y}/G_{\xi})(\tilde y-\tilde y_b)}{\tilde t-\tilde t_b}, 
\ea
\eeq
and describes the plane tangent to the wave at the inflection point (see Fig.2). In addition, 
the gradient of $V$ behaves like
\beq
\nabla_{(\tilde x,\tilde y)}V\sim \frac{\nabla_{(\xi,\tilde y)}G}{G_{\xi}t'+\frac{G_{\xi\xi\xi}}{2|G_{\xi}|}
\left(\frac{\tilde x'+(G_{\tilde y}/G_{\xi})\tilde y'-Gt'}{\tilde t'}+
\frac{G_{\xi\xi\tilde y}}{G_{\xi\xi\xi}}\tilde y'\right)^2+
\frac{\alpha}{2|G_{\xi}|G_{\xi\xi\xi}}\tilde {y'}^2 },
\eeq
implying that, in the narrow strip (\ref{strip_Hopf1})-(\ref{strip_Hopf2}),
\beq
\nabla_{(\tilde x,\tilde y)}V\sim \frac{1}{\tilde t'}\left(1,\frac{G_{\tilde y}}{G_{\xi}}\right)
\eeq
while outside, in the region $|X|=O(|\epsilon |)$, it becomes finite. 
We finally remark that the tangent plane described by (\ref{simil_asymp_1}) passes through the breaking point. 

\subsubsection{At breaking} 

As $\tilde t \uparrow \tilde t_b$, 
the inflection point becomes the breaking point: $\tilde{\vec x}_f\to \tilde{\vec x}_b$, 
the above tangent plane (now tangent at the breaking point) becomes vertical, with equation 
$\tilde {\vec k}\cdot (\tilde{\vec x}-\tilde{\vec x}_b)=0$,       
the above strip reduces to the breaking point $\tilde{\vec x}_b$, and the slopes 
$V_{\tilde x},V_{\tilde y}\to\infty$ as $(\tilde t -\tilde t_b)^{-1}$ (see Fig.3). 

At the breaking time $\tilde t=\tilde t_b$, one can give an explicit description of the 
vertical inflection in the following two subcases.

\vskip 5pt
\noindent
i) $\tilde y'=0$. In this case, the cubic (\ref{cubic}) simplifies to ${\xi'}^3=\gamma \tilde x'$, 
and the solution $V$ exhibits the typical vertical inflection preceeding the wave breaking:
\beq
V\sim G\left(\xi_b+\sqrt[3]{\gamma(\tilde x-\tilde x_b)},\tilde y_b \right)~\Rightarrow~
V_{\tilde x}\sim \frac{1}{3}\sqrt[3]{\frac{6|G_{\xi}|}{G_{\xi\xi\xi}}}
\frac{G_{\xi}}{\sqrt[3]{(\tilde x-\tilde x_b)^2}}.
\eeq

\vskip 5pt
\noindent
ii) $\tilde x'=0$. In this case the cubic simplifies to 
${\xi'}^3-\gamma(G_{\tilde y}/G_{\xi})\tilde y'\sim 0$,  
and $V$ describes a vertical inflection also wrt $\tilde y$:
\beq
V\sim 
G\left(\xi_b+\sqrt[3]{\gamma\frac{G_{\tilde y}}{G_{\xi}}(\tilde y-\tilde y_b)},\tilde y_b+\tilde y'\right)~
\Rightarrow~
V_{\tilde y}\sim \frac{1}{3}\sqrt[3]{\frac{6G_{\tilde y}}
{G_{\xi\xi\xi}}}\frac{G_{\xi}}{\sqrt[3]{{(\tilde y-\tilde y_b)}^2}}.
\eeq

\subsubsection{After breaking}

After breaking, the solution becomes three-valued in a domain of the $(\tilde x,\tilde y)$ - plane, 
and does not describe any physics;   
nevertheless a detailed study of the multivalued region is important, in view of a proper regularization 
of the model, and/or in view of the introduction of a proper single-valued shock replacing 
the multivalued solution. 

If $\tilde t>\tilde t_b$ $(\tilde t'>0)$, in the regime (\ref{max_balance_Hopf}), the SM equation ${\cal S}=0$:  
\beq
\label{ellipse_Hopf}
G_{\xi\xi\xi}{\xi'}^2+2G_{\xi\xi\tilde y}{\xi'}\tilde y'+G_{\xi\tilde y\tilde y}\tilde {y'}^2=|G_{\xi}|\epsilon
\eeq
describes an elliptic paraboloid in the ($\xi,\tilde y,\tilde t$) space, with minimum at 
the point $(\vec\xi_b,\tilde t_b)$ (see Fig.4). 
The intersection of the elliptic paraboloid with any $\tilde t$-constant plane, $\tilde t>\tilde t_b$, 
defines an ellipse in the ($\xi,\tilde y$) - plane 
(see Fig.4), whose coordinates vary in the following intervals of order $O(|\epsilon|^{1/2})$: 
\beq
\label{ellipse_2}
\ba{l}
|\tilde y'|\le \sqrt{\frac{|G_{\xi}|G_{\xi\xi\xi}\epsilon}{\alpha}},~~~~
|{\xi'}|\le \sqrt{\frac{|G_{\xi}|G_{\xi\tilde y\tilde y}\epsilon}{\alpha}}.
\ea
\eeq

Eliminating $\xi'$ from equations (\ref{ellipse_Hopf}) and (\ref{cubic}), one obtains the SM equation in space-time 
coordinates:
\beq
\label{Delta=0}
\ba{l}
\Big[ 3|G_{\xi}|G^2_{\xi\xi\xi}\left(\tilde x'+\frac{G_{\tilde y}}{G_{\xi}}\tilde y'-G\tilde t'+
\frac{G_{\tilde y\tilde y}}{2G_{\xi}}\tilde {y'}^2-G_{\tilde y}\tilde y'\tilde t'+
\frac{G_{\tilde y\tilde y\tilde y}}{6G_{\xi}}\tilde {y'}^3 \right)+   \\
\frac{G_{\xi\xi\xi}G_{\xi\xi\tilde y}G_{\xi\tilde y\tilde y}}{2} 
\left(\frac{G_{\xi}}{G_{\xi\tilde y\tilde y}}\epsilon+{y'}^2\right) \tilde y'-
\alpha G_{\xi\xi\tilde y} \left(\frac{|G_{\xi}|G_{\xi\xi\xi}}{\alpha}\epsilon-\tilde {y'}^2\right)\tilde y' \Big]^2=\\
\alpha^3 \Big(\frac{|G_{\xi}|G_{\xi\xi\xi}}{\alpha}\epsilon -\tilde {y'}^2 \Big)^3,
\ea
\eeq
coinciding with the $\Delta =0$ condition. It describes a closed curve of the $(\tilde x,\tilde y)$ plane 
possessing two cusps at the points
\beq\label{cusps_Hopf}
\tilde{\vec x}^{\pm}_c(\tilde t')\sim \tilde{\vec x}_b 
\mp \sqrt{\frac{|G_{\xi}|G_{\xi\xi\xi}\epsilon}{\alpha}}\left(\frac{G_{\tilde y}}{G_{\xi}}, 1 \right).
\eeq

If ($x_{\pm}(\tilde y),\tilde y$) are the two intersections of this curve with the horizontal line    
$\tilde y=$ const, then 
\beq
\ba{l}
\tilde x_+(\tilde y)-\tilde x_-(\tilde y)=-\frac{4}{\gamma}|Q|^{3/2}=O(|\epsilon|^{3/2}).
\ea
\eeq
It follows that this curve is the boundary of a narrow region of thickness O($\epsilon^{3/2}$)  
in the longitudinal direction, and of thickness O($\epsilon^{1/2}$)   
in the transversal direction. In Fig.5 we show its time evolution after breaking.

Away from the cusps, 
the solution of the Hopf equation is three valued, but two of the branches coincide:
\beq
V_0=G(\xi_b+\xi'_0,\tilde y),~~V_+=V_-=G(\xi_b+\xi'_+,\tilde y)
\eeq 
and the slopes of the coincident solutions are $\infty$. We remark that the closed curve, 
being the envelope of the 
intersections with the $(\tilde x,\tilde y)$ - plane of these vertical planes tangent to the wave, 
is a caustic (see Fig.6). 
 
At the two cusps, characterized by 
the condition $Q=R=0$, the three real roots of the cubic coincide and the corresponding three 
solutions of the Hopf equation coincide as well:
\beq
V_0=V_+=V_-=G\left(\xi_b\pm \frac{G_{\xi\xi\tilde y}|G_{\xi}|}{G_{\xi\xi\xi}}\sqrt{\frac{2G_{\xi\xi\xi}t'}{\alpha}},
\tilde y_b\pm |G_{\xi}|\sqrt{\frac{2G_{\xi\xi\xi}t'}{\alpha}} \right).
\eeq

Inside the caustic, the discriminant $\Delta$ is strictly negative, the cubic admits 
three different real roots and the solution of the Hopf equation is three-valued. This is the  
multivalued region  that has to be replaced by a proper shock layer, 
whose features depend on the desired regularization. 
Outside the singularity manifold, $\Delta>0$ and the solution of the Hopf equation is single valued.  

We end this section remarking that the similarity solution before breaking, 
the vertical inflection at breaking, and the caustic after breaking 
make clear the universal character of the gradient catastrophe of two-dimensional waves evolving 
according to the Hopf equation. As we shall see in the following sections, similar considerations hold 
for the gradient catastrophe of two-dimensional waves evolving according to the dKP equation. 
Similar results on a different model can be found in \cite{Kuznetsov}. A general singularity theory 
for caustics and wave fronts can be found in \cite{Arnold}.  

\subsection{Longtime breaking of dKP waves}

Inverting the transformation (\ref{change_variables}), the formulas of \S 5.1 allow one 
to describe the longtime breaking of dKP solutions. Now
\beq\label{U}
\ba{c}
U(x,y,t)=\frac{1}{\sqrt{t}}G(\xi,\tilde y),  \\
\xi=x+\frac{y^2}{4t}-2\sqrt{t}G(\xi,\tilde y),~~~\tilde y=\frac{y}{2t}
\ea
\eeq
and 
\beq\label{grad_U}
\nabla_{(x,y)}U=\frac{1}{\sqrt{t}}
\frac{\left(G_{\xi}(\xi,\tilde y),\frac{y}{2t}G_{\xi}(\xi,\tilde y)+\frac{1}{2t}G_{\tilde y}(\xi,\tilde y) \right)}
{1+2\sqrt{t}G(\xi,\tilde y)}.
\eeq

Let $\vec \xi_b=(\xi_b,\tilde y_b)$ be the breaking parameters and $(\tilde x_b,\tilde y_b,\tilde t_b)$ 
be the breaking point associated with the solution (\ref{sol_Hopf}) of the 
Hopf equation (\ref{Hopf}), and characterized by 
equations (\ref{tilde_t_b}) - (\ref{tilde_x_b}). Then a localized solution $U(x,y,t)$ of equation  
(\ref{U}) first breaks at the time 
\beq
\label{t_b}
t_b=\left(\frac{\tilde t_b}{2}\right)^2=(2G_{\xi})^{-2},
\eeq 
in the point $\vec x_b=(x_b,y_b)$ of the $(x,y)$ - plane given by
\beq
\label{x_b}
x_b=\tilde x_b-\tilde y^2_b t_b,~~~~y_b= 2\tilde y_bt_b;
\eeq
we remark that $\vec x_b$ is the intersection of the parabola $x+y^2/4t_b=\tilde x_b$ with the straight line $y=y_b$. 

The inflection point $\vec x_f(t)=(x_f(t),y_f(t))$ is given by
\beq
\ba{l}
x_f(t)=x_b+2G(\sqrt{t}-\sqrt{t_b})-\tilde y^2_b(t-t_b) \sim  x_b+ \\
\left[2|G_{\xi}|G(1-\epsilon/4)-\tilde y^2_b\right](t-t_b),   \\
y_f(t)=y_b+2\tilde y_b(t-t_b), 
\ea
\eeq
where the small parameter $\epsilon$, introduced in (\ref{epsilon}),  reads 
\beq
\epsilon=\frac{t-t_b}{t_b}
\eeq
in terms of the dKP coordinates. The inflection point $\vec x_f(t)$ 
is the intersection of the parabola $x+y^2/4t=\tilde x_f=\tilde x_b+2G(\sqrt{t}-\sqrt{t_b})$ with the 
straight line $y=y_f(t)$ (see Fig.7), where $\tilde x_f$ is defined in (\ref{tilde_inflection}) . 
At the inflection point, $U_{xx}=U_{xy}=0$, and $U$ and its gradient take the following values:
\beq
\ba{l}
U=\frac{G}{\sqrt{t}},~~\nabla_{(x,y)}U=\sqrt{\frac{t_b}{t}}\frac{1}{t-t_b}\left(1,\tilde y_b+
\frac{1}{2t}\frac{G_{\tilde y}}{G_{\xi}} \right) \sim  
\frac{1}{t-t_b}\left(1,\tilde y_b+2G_{\tilde y}G_{\xi}\right).
\ea
\eeq

Throughout this section we make a systematic use of the expressions of the small variables 
$\tilde x',\tilde y',\tilde t'$ in terms of the dKP variables:
\beq\label{tildey}
\ba{l}
\tilde x'=(x-x_b)+\tilde y_b\left( (y-y_b)-\tilde y_b(t-t_b)+\frac{(y-y_b)^2}{2y_b}\right)(1-\epsilon 
+O(\epsilon^2)),\\
\tilde y'=\tilde y_b\frac{y-y_b-2\tilde y_b(t-t_b)}{y_b}(1-\epsilon +O(\epsilon^2))=
\tilde y_b\frac{y-y_f(t)}{y_b}(1-\epsilon +O(\epsilon^2)),  \\
\tilde t'= \frac{t-t_b}{\sqrt{t_b}}(1-\epsilon/4).
\ea
\eeq

\subsubsection{Before breaking}

If $t<t_b$, in the regime  (\ref{max_balance_Hopf}), 
the real solution $U$ of (\ref{U}) is single valued and described by 
\beq
U(x,y,t)\sim \frac{1}{\sqrt{t}}G(\xi_b+\xi'_0,\tilde y_b+\tilde y')
\eeq
where $\xi'_0$ is defined in (\ref{sol_cubic1}) by Cardano's formula, and $\tilde x',\tilde y'$ are given in 
(\ref{tildey}).

To obtain a more explicit solution, we restrict the analysis to the narrow strip around the 
inflection point, defined by the conditions
\beq
\label{strip_long}
\ba{l}
|X|\sim |\vec k \cdot (\vec x-\vec x_f(t))|=O(|\epsilon|^{p+1}), ~~~ \frac{|y-y_f(t)|}{t_b}=O(|\epsilon|^{q}),
\ea
\eeq 
where $p,q$ are defined in (\ref{strip_Hopf2}) and the vector 
\beq
\vec k =(1,\tilde y_b+2G_{\xi}G_{\tilde y})
\eeq
characterizes the breaking direction. On this strip the cubic linearizes and its solution reads 
\beq\label{xi}
\xi'=\frac{(x-x_b)-(2G|G_{\xi}| -\tilde y^2_b)(t-t_b)+
(\tilde y_b+2G_{\xi}G_{\tilde y})(y-y_b-2\tilde y_b(t-t_b))}{\frac{G_{\xi}}{\sqrt{t_b}}(t-t_b)},
\eeq 
and $U$ is approximated by the exact similarity solution of equation (\ref{U_equ}):
\beq
\label{U_sim}
U \sim \frac{x+\frac{y^2}{4t}-
\tilde x_b+\frac{G_{\eta}}{G_{\xi}}(\frac{y}{2t}-\tilde y_b)}{2\sqrt{t}(\sqrt{t}-\sqrt{t_b})}.
\eeq 
Equation (\ref{U_sim}) is equivalent, at the same order, to 
\beq\label{U_sim_2}
\ba{l}
U \sim \sqrt{\frac{t_b}{t}}\frac{x-x_b+\eta^2_b(t-t_b)+(\eta_b+2G_{\xi}G_{\eta})[y-y_b-2\eta_b(t-t_b)]}{t-t_b} \sim \\
\frac{x-x_b+\eta^2_b(t-t_b)+(\eta_b+2G_{\xi}G_{\eta})[y-y_b-2\eta_b(t-t_b)]}{t-t_b},
\ea
\eeq
describing the plane tangent to the wave at the point $\vec x_f$ (see Fig.7). The gradient of $U$ reads, 
from (\ref{grad_U}),
\beq
\nabla_{(x,y)}U=\frac{1}{\sqrt{t}}
\frac{ \left( G_{\xi}, \frac{y}{2t}G_{\xi}+\frac{1}{2t} G_{\tilde y} \right) }
{\frac{G_{\xi}}{\sqrt{t_b}}t'+ 
\frac{G_{\xi\xi\xi}}{2|G_{\xi}|}\left[ \left(\xi'+\frac{G_{\xi\xi\tilde y}}{G_{\xi\xi\xi}}\tilde y'\right)^2 +
\frac{\alpha}{G^2_{\xi\xi\xi}}\tilde {y'}^2 \right]},
\eeq
where $\xi'$ and $\tilde y'$ are given in (\ref{xi}) and (\ref{tildey}). On the narrow strip (\ref{strip_long}) it 
is very large: 
\beq
\nabla_{(x,y)}U\sim \frac{1}{t-t_b}\left(1,\tilde y_b+2G_{\tilde y}G_{\xi}\right),
\eeq 
while outside, for $|X|=O(|\epsilon |)$, it is finite.

\subsubsection{At breaking}

As $t\uparrow t_b$, 
the inflection point becomes the breaking point: ${\vec x}_f\to {\vec x}_b$, 
the above tangent plane becomes vertical, with equation ${\vec k}\cdot ({\vec x}-{\vec x}_b)$ and  
$U_{x},U_{y}\to\infty$ as $(t -t_b)^{-1}$. 

At the breaking time $t=t_b$, one can give an explicit description of the 
vertical inflection in the following two subcases.

If $y=y_b$, the solution exhibits the typical vertical inflection 
preceeding the wave breaking:
\beq
U\sim \frac{1}{\sqrt{t_b}}G\left(\xi_b+\sqrt[3]{\gamma(x-x_b)},\tilde y_b \right)~\Rightarrow~
U_{x}\sim \frac{\sqrt[3]{\gamma}}{3\sqrt{t_b}}\frac{G_{\xi}}{\sqrt[3]{(x-x_b)^2}}.
\eeq

If $x-x_b+\eta_b(y-y_b)+\frac{(y-y_b)^2}{4t_b}=0$ (a line approximately tangent to the parabola 
at ${\vec x}_b$), $U$ describes a vertical inflection also wrt $y$:
\beq
U\sim \frac{1}{\sqrt{t_b}}G\left(\xi_b+\sqrt[3]{2\gamma G_{\xi}G_{\tilde y}(y-y_b)},
\frac{y}{2t_b} \right)~\Rightarrow~U_{y}\sim 
\frac{\sqrt[3]{2\gamma G_{\xi}G_{\eta}}}{3\sqrt{t_b}}\frac{G_{\xi}}{\sqrt[3]{{(y-y_b)}^2}}.
\eeq

\subsubsection{After breaking}

If $t>t_b$, in the regime (\ref{max_balance_Hopf}), the intersection of the SM 
with any $t$ - constant plane, defines an ellipse in the ($\xi,\eta$) - plane 
(see Fig.4), corresponding to the following caustic in the ($x,y$) - plane, defined, as in (\ref{Delta=0}), by 
\beq
\label{Delta=0_long}
\ba{l}
\Big[ 3|G_{\xi}|G^2_{\xi\xi\xi} X + 
\frac{G_{\xi\xi\xi}G_{\xi\xi\tilde y}G_{\xi\tilde y\tilde y}}{2} 
\left(\frac{G_{\xi}}{G_{\xi\tilde y\tilde y}}\epsilon + \tilde {y'}^2\right) \tilde {y'}- \\
\alpha G_{\xi\xi\tilde y} \left(\frac{|G_{\xi}|G_{\xi\xi\xi}}{\alpha}\epsilon-\tilde {y'}^2\right) \tilde {y'} \Big]^2=
\alpha^3 \Big(\frac{|G_{\xi}|G_{\xi\xi\xi}}{\alpha}\epsilon -\tilde {y'}^2 \Big)^3,
\ea
\eeq
where now
\beq
\ba{l}
\tilde {y'}=\tilde y_b \frac{y-y_f(t)}{y_b}, \\
X= x-x_b - [2G|G_{\xi}|(1-\frac{1}{4}\epsilon)-\tilde y^2_b (1-\epsilon )](t-t_b) + \\
\left[ (\tilde y_b+2G_{\xi}G_{\tilde y})(y-y_f(t))+ G^2_{\xi}(y-y_b)^2 \right](1-\epsilon ) +
\frac{G_{\tilde y\tilde y}}{2G_{\xi}}\tilde y^2_b\left(\frac{y-y_f(t)}{y_b}\right)^2 - \\
2|G_{\xi}|G_{\tilde y}\tilde y_b(t-t_b)\frac{y-y_f(t)}{y_b}+
\frac{G_{\tilde y\tilde y\tilde y}}{6G_{\xi}}\tilde y^3_b\left(\frac{y-y_f(t)}{y_b}\right)^3.
\ea
\eeq
The caustic exhibits two cusps at the points
\beq\label{cusps_long}
\vec x^{\pm}(t) \sim  \vec x_b  \mp \sqrt{2\frac{G_{\xi\xi\xi}}{\alpha}(t-t_b)}.
\left(\tilde y_b+2G_{\xi}G_{\eta},-1\right)
\eeq
In addition, if $\left( x^{\pm}(y),y \right)$ are the two intersection points of the caustic with 
the line $y=$ const., we have
\beq
x^+(y)-x^-(y)=\frac{2\alpha^{3/2}}{3|G_{\xi}|G_{\xi\xi\xi}}\Big(\frac{|G_{\xi}|G_{\xi\xi\xi}}{\alpha}\epsilon-
\left(\frac{y-y_f}{2t_b}\right)^2 \Big)^{3/2}=O(|\epsilon |^{3/2}); 
\eeq 
therefore the caustic is the boundary of a narrow region of thickness O($|\epsilon |^{3/2}$)  
in the longitudinal direction, and of thickness O($\sqrt{2\frac{G_{\xi\xi\xi}}{\alpha}(t-t_b)}$)    
in the transversal direction (see Fig.5). On it, the discriminant $\Delta$ of the cubic is zero and, 
away from the cusps, 
the solution of equation (\ref{U}) is three valued, two of the branches coincide:
\beq
U_0=\frac{1}{\sqrt{t}}G(\xi_b+\xi'_0,\tilde y_b+\tilde y'),~~U_+=U_-=\frac{1}{\sqrt{t}}G(\xi_b+\xi'_+,\tilde y_b+\tilde y')
\eeq 
and the slopes of the coincident solutions are $\infty$ (see Fig.6). At the two cusps, characterized by 
the condition $Q=R=0$, the three real roots of the cubic coincide and the corresponding three 
solutions of equation (\ref{U}) coincide as well:
\beq
\ba{l}
U_0=U_+=U_-=\frac{1}{\sqrt{t}}
G\Big( \xi_b\pm 2\frac{G_{\xi\xi\eta}|G_{\xi}|^{3/2}}{G_{\xi\xi\xi}}\sqrt{\frac{G_{\xi\xi\xi}}{\alpha}(t-t_b)},\\
\tilde y_b \mp 2|G_{\xi}|^{3/2}\sqrt{\frac{G_{\xi\xi\xi}}{\alpha}(t-t_b)} \Big) .
\ea
\eeq
Inside the caustic, the discriminant $\Delta$ is strictly negative, the cubic admits 
three different real roots and the solution of equation  (\ref{U}) is three-valued. 
Outside, $\Delta>0$ and the solution of the equation (\ref{U}) is single valued.  

The formulae of this section describe, after replacing $U$ by $u$, the longtime breaking 
of the dKP solutions $u$ if, for instance, the dKP initial data $u_0(x,y)=u(x,y,0)$ are small. For 
small initial data, the inverse spectral transform for dKP, developed in \cite{MS3}, simplifies 
enormously. The  RH spectral data are expressed in terms of the initial data as follows:
\beq
R_2(\zeta_1,\zeta_2)
\sim \frac{1}{\pi i}\int\limits_{\RR^2}\frac{d\xi' dy}{\zeta_1-\xi'}u_{0\xi'}(\xi'+\zeta_2 y,y),
\eeq   
and function $G$, appearing in all formulas of this section,  
is also given explicitly in terms of $u_0$ via (\ref{asympt_dKP_2}): 
\beq
G(\xi,\eta)\sim 
\frac{1}{2\pi^2}\int\limits_{\RR^3}\frac{d\xi'd\mu dy}{\xi-\mu^2-\xi'}u_{0\xi'}(\xi'-\eta y,y).
\eeq 

Summarizing, we have shown that small and localized initial data evolving according to the 
dKP equation break in the longtime regime, and we have described the analytic aspects of such 
breaking in a surprisingly explicit way.     
The similarity solution before breaking, the vertical inflection at breaking, and the caustic  
after breaking make clear the universal character of such a gradient catastrophe.

\section{The gradient catastrophe at finite time}

Small and localized initial waves evolving according to the 
dKP equation break in the longtime regime in the way described in \S 5.2. If the initial wave is not small, 
the gradient catastrophe 
takes place at a finite time and, in this section, we study the universal features of 
this phenomenon immediately before breaking.

The analysis is very similar to that of \S 5; the main difference is that, 
while function $G$, appearing in the implicit equation (\ref{sol_Hopf}) for the longtime solution 
of dKP, is an essentially arbitrary smooth function of two variables, function $F$, appearing in 
the implicit equation (\ref{inverse_1}) for the solution of 
dKP at finite time, is a specific function of three variables, connected with the spectral data 
via (\ref{inverse_2}). As we shall see, this implies that, at breaking, $F$ satisfies an extra condition   
(equation (\ref{hidden1}) below) that is instead automatically satisfied when $F$ is replaced by 
its asymptotic form: 
\beq
F(x,y,t) ~\to~ \frac{1}{\sqrt{t}}G\left(x+\frac{y^2}{4t},\frac{y}{2t}\right).
\eeq     
     
The inverse problem of the dKP equation, 
summarized in formulae (\ref{inverse_1}) and (\ref{inverse_2}), defines an  
implicit system of equations for the solution $u$:
\beq
\label{sol_finite}
\ba{l}
u=F(\zeta,y,t), \\
x=\zeta +2F(\zeta,y,t)t .
\ea
\eeq

In analogy with the considerations of \S 5, the singularity manifold of 
dKP is the two - dimensional manifold characterized by the equation 
\beq\label{SM_1}
{\cal S}(\zeta,y,t)\equiv 1+2F_{\zeta}(\zeta,y,t)t=0,
\eeq
that can be solved with respect to $t$, if $F_{\zeta}+F_{\zeta t}t\ne 0$, giving 
\beq\label{SM2}
t=\check t(\zeta,y).
\eeq 
Since 
\beq
\label{gradu}
\nabla_{(x,y)}u=\frac{\nabla_{(\zeta,y)}F(\zeta,y,t)}{1+2F_{\zeta}(\zeta,y,t)t},
\eeq
the gradient of the localized wave becomes infinity on the SM, and the wave ``breaks''. 
Let $t_b$ the first time at which $\cal S=0$ in a point $\vec\zeta_b=(\zeta_b,y_b)$ of the 
($\zeta,y$) - plane:     
\beq\label{t_breaking}
1+2F_{\xi}(\vec\zeta_b,t_b)t_b=0~~~\Rightarrow~~~t_b=\check t(\vec\zeta_b);
\eeq
then we obtain the following conditions characterizing the breaking point ($\vec\zeta_b,t_b$):
\beq
\label{summa}
\ba{l}
1+2t_bF_{\zeta}(\vec\zeta_b,t_b)=0                                                         \\
F_{\zeta}(\vec\zeta_b,t_b)<0, ~~~F_{\zeta}(\vec\zeta_b,t_b)+t_bF_{\zeta t}(\vec\zeta_b,t_b)<0,     \\
F_{\zeta\zeta}(\vec\zeta_b,t_b)=F_{\zeta\ y}(\vec\zeta_b,t_b)=0, \\
F_{\zeta\zeta\zeta}(\vec\zeta_b,t_b)>0,~~~
\beta \equiv F_{\zeta\zeta\zeta}(\vec\zeta_b,t_b)F_{\zeta yy}(\vec\zeta_b,t_b)-F^2_{\zeta\zeta y}(\vec\zeta_b,t_b)>0.
\ea
\eeq
Due to (\ref{sol_finite}b), at the breaking time $t_b$ the wave breaks 
in the point $\vec x_b=(x_b,y_b)$ of the $(x,y)$ - plane defined by
\beq\label{breaking_place}
x_b=\zeta_b+2 F(\vec\zeta_b,t_b) t_b.
\eeq
As before, we evaluate equations (\ref{sol_finite}b) and (\ref{SM_1}) near breaking, in the regime:
\beq
\label{near_breaking_2}
x=x_b+x',~~y=y_b+y',~~t=t_b+t',~~\zeta=\zeta_b+{\zeta'},
\eeq
where $x',y',t',{\zeta'}$ are small, obtaining, at the leading order, 
the cubic 
\beq
{\zeta'}^3+a(y'){\zeta'}^2+b(y',t'){\zeta'}-\gamma X(x',y',t')=0,
\eeq
where now
\beq
\label{ab}
\ba{l}
a(y')=\frac{3F_{\zeta\zeta y}}{F_{\zeta\zeta\zeta}}y',~~~~~
b(y',t')=3 [\frac{2(F_{\zeta}+t_bF_{\zeta t})}{F_{\zeta\zeta\zeta}}\epsilon+
\frac{F_{\zeta yy}}{F_{\zeta\zeta\zeta}}{y'}^2],  \\
X(x',y',t')=x'-2F(\zeta_b,y,t) t'-2\left[F(\zeta_b,y,t)-F\right]t_b\sim  \\
x'+\frac{F_{y}}{F_{\zeta}}y'-2(F+t_bF_t)t'-\frac{F_{yy}}{2|F_{\zeta}|}{y'}^2-
2(F_{y}+t_bF_{y t})y't'-\frac{F_{yyy}}{6|F_{\zeta}|}{y'}^3,     \\
\gamma=\frac{6|F_{\zeta}|}{F_{\zeta\zeta\zeta}},
\ea
\eeq
and $\epsilon$ is given again by (\ref{epsilon}),  
corresponding to the maximal balance $|\zeta'|,|y'|=O(|\epsilon|^{1/2})$ and $X=O(|\epsilon|^{3/2})$. 
In formula (\ref{ab}), and in the rest of 
this section, all partial derivatives of $F$ whose 
arguments are not indicated are meant to be evaluated at the breaking point ($\vec\zeta_b,t_b$).

The three roots of this cubic are given by Cardano's formulas
(\ref{sol_cubic1})-(\ref{discriminant}), where now
\beq
\label{def_QR}
\ba{l} 
Q(y',t')= \frac{2}{F_{\zeta\zeta\zeta}}(F_{\zeta}+t_bF_{\zeta t})\epsilon +
\frac{\beta}{F^2_{\zeta\zeta\zeta}}{y'}^2 .
\ea
\eeq
Function $\cal S$ reads, at the leading order,
\beq
\label{SM_2}
{\cal S}\sim 2(F_{\zeta}+t_bF_{\zeta t})t'+
\left(F_{\zeta\zeta\zeta}{\zeta'}^2+2F_{\zeta\zeta y}y'{\zeta'}+F_{\zeta yy}{y'}^2\right)t_b.
\eeq

For $t<t_b~(t'<0)$, since the coefficient $Q$ is strictly positive (see (\ref{summa}) and 
(\ref{def_QR})), so is the 
discriminant $\Delta$ in (\ref{discriminant}). It follows that $\xi'_0$ is real and $\xi'_{\pm}$ 
are complex conjugate roots; therefore the solution of dKP is single valued and it is described Cardano's 
formula. In addition, function ${\cal S}$ is also strictly positive, 
implying that the gradient of $u$ is regular near breaking, in the space-time region (\ref{max_balance_Hopf}). 

Restricting the asymptotic region in the $(x,y)$ - plane to the narrow strip $|y'|=O(|\epsilon|^q),
~|X|=O(|\epsilon|^{p+1})$, 
where $p,q$ are defined by the inequalities (\ref{strip_Hopf2}),  
then $|{\zeta'}|=O(|\epsilon|^p)$, ${\zeta'}^3,a{\zeta'}^2<<b{\zeta'}\sim \gamma X$, and 
the expressions for $X$ and ${\zeta'}$ simplify: 
\beq
\ba{l}
X=x'+\frac{F_{y}}{F_{\zeta}}y'-2(F+t_b F_t)t'+O(|\epsilon |^{p+1}),     \\
~~~~   \\
{\zeta'}\sim \frac{x'+\frac{F_{y}}{F_{\zeta}}y'-2(F+t_b F_t)t'}{2(F_{\zeta}+t_b F_{\zeta t})t'},
\ea
\eeq
implying 
\beq\label{sol_balance_2}
\ba{l}
u\sim F + \frac{F_{\zeta}}{2(F_{\zeta}+t_b F_{\zeta t})}\frac{x'+(F_{y}/F_{\zeta})y'}{t'}.
\ea
\eeq
Since 
\beq\label{similarity1}
u_{sim}\equiv \frac{(x-x_b)+(F_{y}/F_{\zeta})(y-y_b)+ c (t-t_b)}{t-t_b}
\eeq
is an exact similarity solution of dKP, where $c$ is an arbitrary constant, 
comparing (\ref{sol_balance_2}) and (\ref{similarity1}),  
it follows that $F$ must satisfy the following condition at breaking:
\beq\label{hidden1}
\ba{l}
F_{\zeta}+2t_b F_{\zeta t}=0 \;\;\;\left(\Rightarrow\;F_{\zeta t}=F^2_{\zeta}\right),
\ea
\eeq
implying that
\beq
\label{SM_3}
\ba{l}
{\cal S}\sim F_{\zeta}t'+
\left(F_{\zeta\zeta\zeta}{\zeta'}^2+2F_{\zeta\zeta y}y'{\zeta'}+F_{\zeta yy}{y'}^2\right)t_b, \\
Q(y',t')=-\frac{2}{F_{\zeta\zeta\zeta}}\epsilon +\frac{\beta }{F^2_{\zeta\zeta\zeta}}{y'}^2, ~~~~
{\zeta'}\sim \frac{x'+(F_{y}/F_{\zeta})y'-2(F+t_bF_t)t'}{F_{\zeta}t'}.
\ea
\eeq 

Therefore, in the narrow strip
\beq\label{strip_finite_t}
|x'+\frac{F_{y}}{F_{\zeta}}y'-2(F+t_b F_t)t'|=O(|\epsilon |^{p+1})
\eeq
of the ($x,y$)-plane, the solution $u$ is described by the exact similarity solution of dKP:
\beq
u \sim u_{sim}=\frac{(x-x_b)+(F_{y}/F_{\zeta})(y-y_b)-(F+2t_bF_t)(t-t_b)}{t-t_b},
\eeq  
while the asymptotic expression for the gradient of $u$ follows from equations (\ref{gradu}) and (\ref{SM_3}):
\beq
\nabla_{(x,y)}u\sim \frac{\nabla_{(\zeta,y)}F}
{F_{\zeta}t'+\frac{F_{\zeta\zeta\zeta}}{2|F_{\zeta}|}\Big(\frac{x'+(F_{y}/F_{\zeta})y'-2(F+t_b F_t)t'}{F_{\zeta}t'} 
+\frac{F_{\zeta\zeta y}}{F_{\zeta\zeta\zeta}}y'\Big)^2+\frac{\alpha}{2|F_{\zeta}|F_{\zeta\zeta\zeta}}{y'}^2}.
\eeq
Then
\beq
\nabla_{(x,y)}u=\frac{1}{t-t_b}(1,F_{y}/F_{\zeta})
\eeq
in the narrow strip (\ref{strip_finite_t}), while outside, for $|X|=O(|\epsilon|)$, 
the gradient is finite. The vector $\vec k=(1,F_{y}/F_{\zeta})$, orthogonal to the strip, 
defines the direction of breaking. 
 
In the limit $t \uparrow t_b$, the narrow strip (\ref{strip_finite_t}) reduces to the 
breaking point,  the above tangent plane becomes vertical, with equation 
$F_{\zeta}(x-x_b)+F_{y}(y-y_b)=0$, and the slopes $u_{x},u_{y}\to\infty$ at that point as $(t -t_b)^{-1}$ 
(see Fig.3). As before, we consider two sections in which the description is simpler.

If $y'=0$, the solution of dKP is described by the typical vertical inflection at the breaking 
point $\vec x_b$:
\beq
u\sim F(\zeta_b+\sqrt[3]{\gamma (x-x_b)},y_b,t_b)~~\Rightarrow~~
u_x\sim \frac{\sqrt[3]{\gamma}}{3}\frac{F_{\zeta}}{\sqrt[3]{(x-x_b)^2}}.
\eeq

If $x'=0$, we have again the vertical inflection at $\vec x_b$:
\beq
u\sim F\left(\zeta_b-\sqrt[3]{\gamma \frac{F_{y}}{|F_{\zeta}|} (y-y_b)},y_b,t_b\right)~\Rightarrow~
u_y\sim -\frac{1}{3}\sqrt[3]{\gamma \frac{F_{y}}{|F_{\zeta}|}}\frac{F_{\zeta}}{\sqrt[3]{(y-y_b)^2}}.
\eeq

\newpage 

 
\begin{center}
\mbox{ \epsfxsize=10cm \epsffile{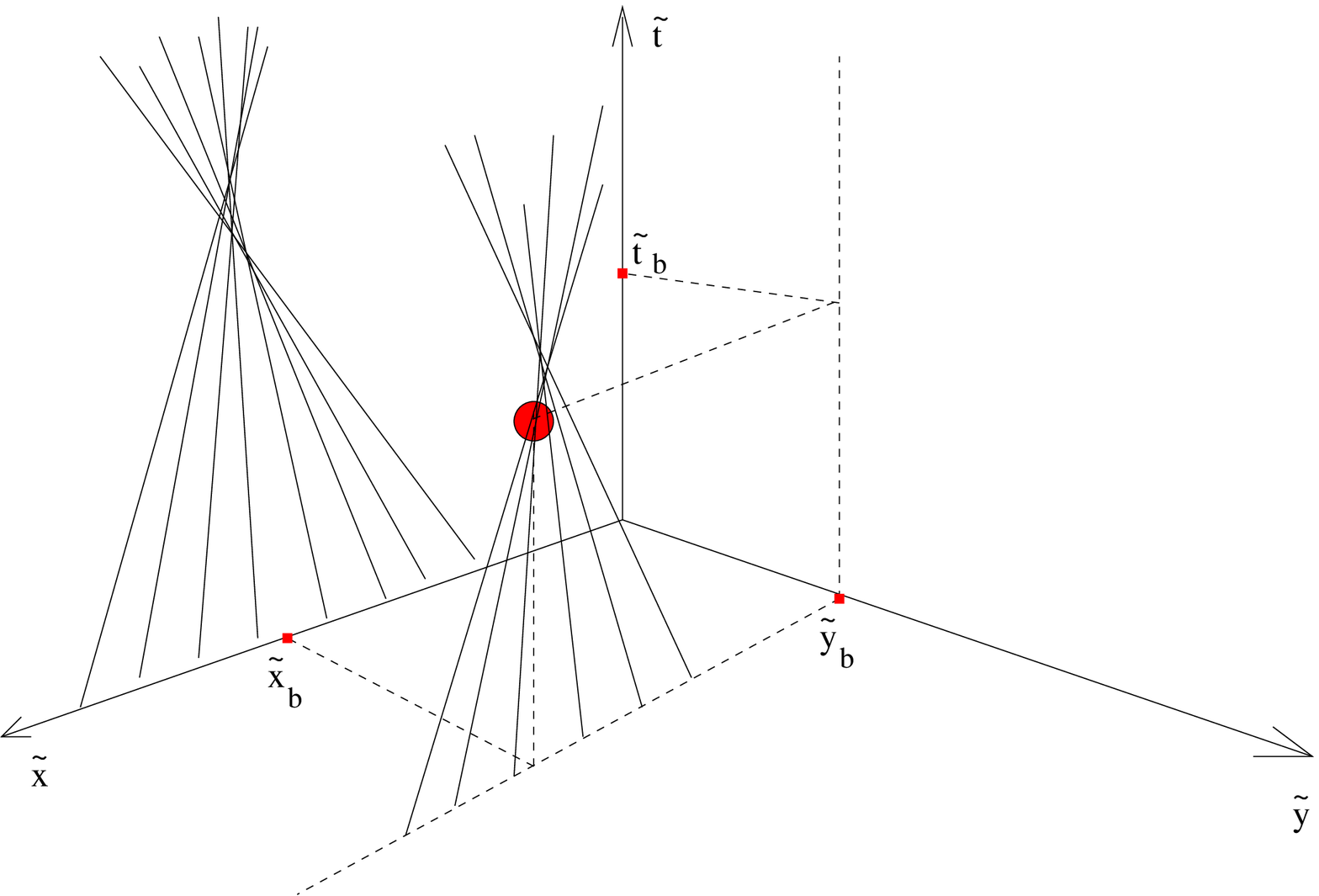}}
\end{center}
Fig.1 \;\;\;\;\;The 2-parameter family of characteristic curves describing the evolution of 2D-waves 
according to the $1+1$ dimensional Hopf equation.


\begin{center}
\mbox{ \epsfxsize=10cm \epsffile{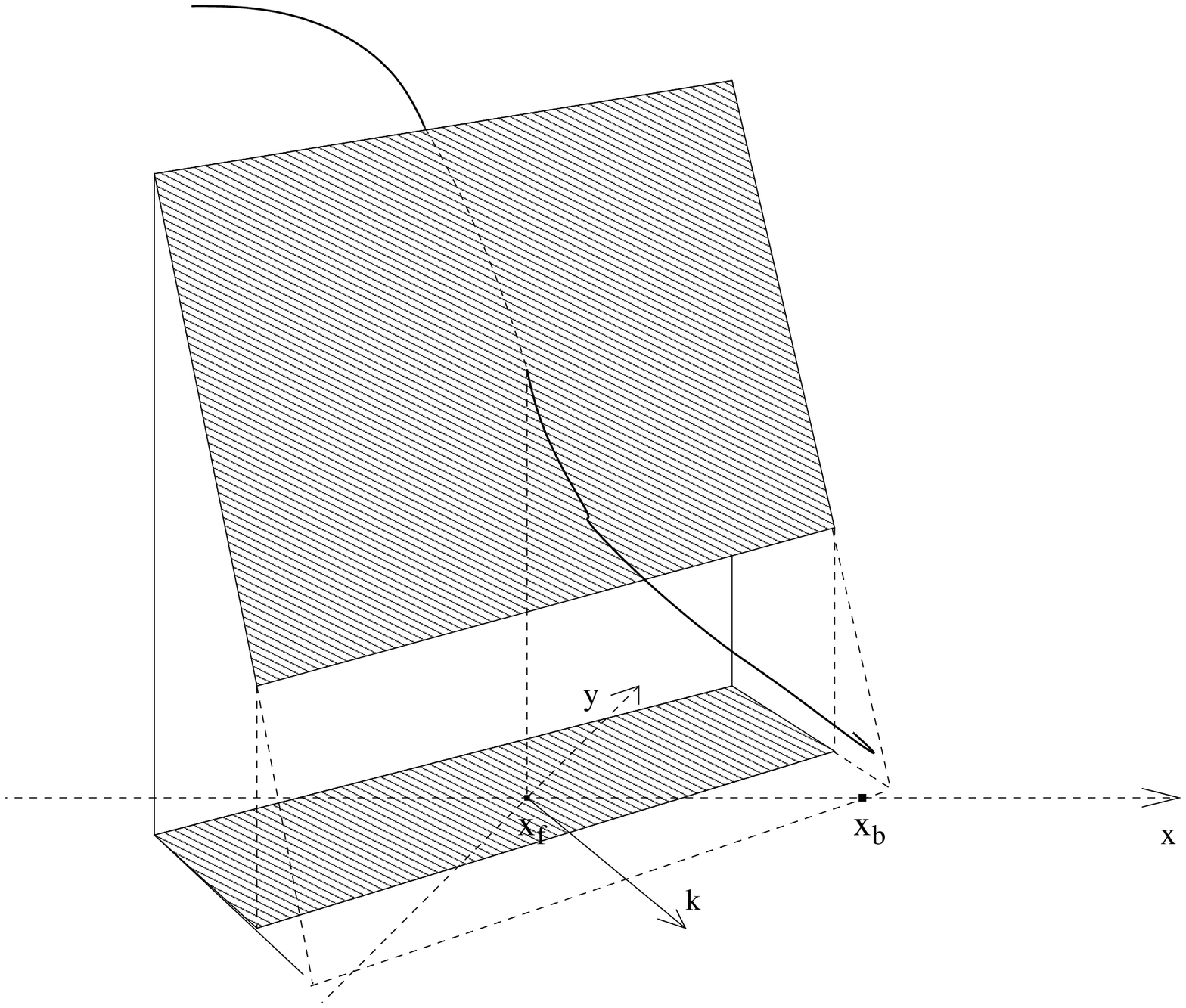}}
\end{center}
Fig.2 \;\;\;\;\;In a narrow strip of the plane, the similarity solution describes 
the plane tangent to the wave, immediately before breaking.


\begin{center}
\mbox{ \epsfxsize=10cm \epsffile{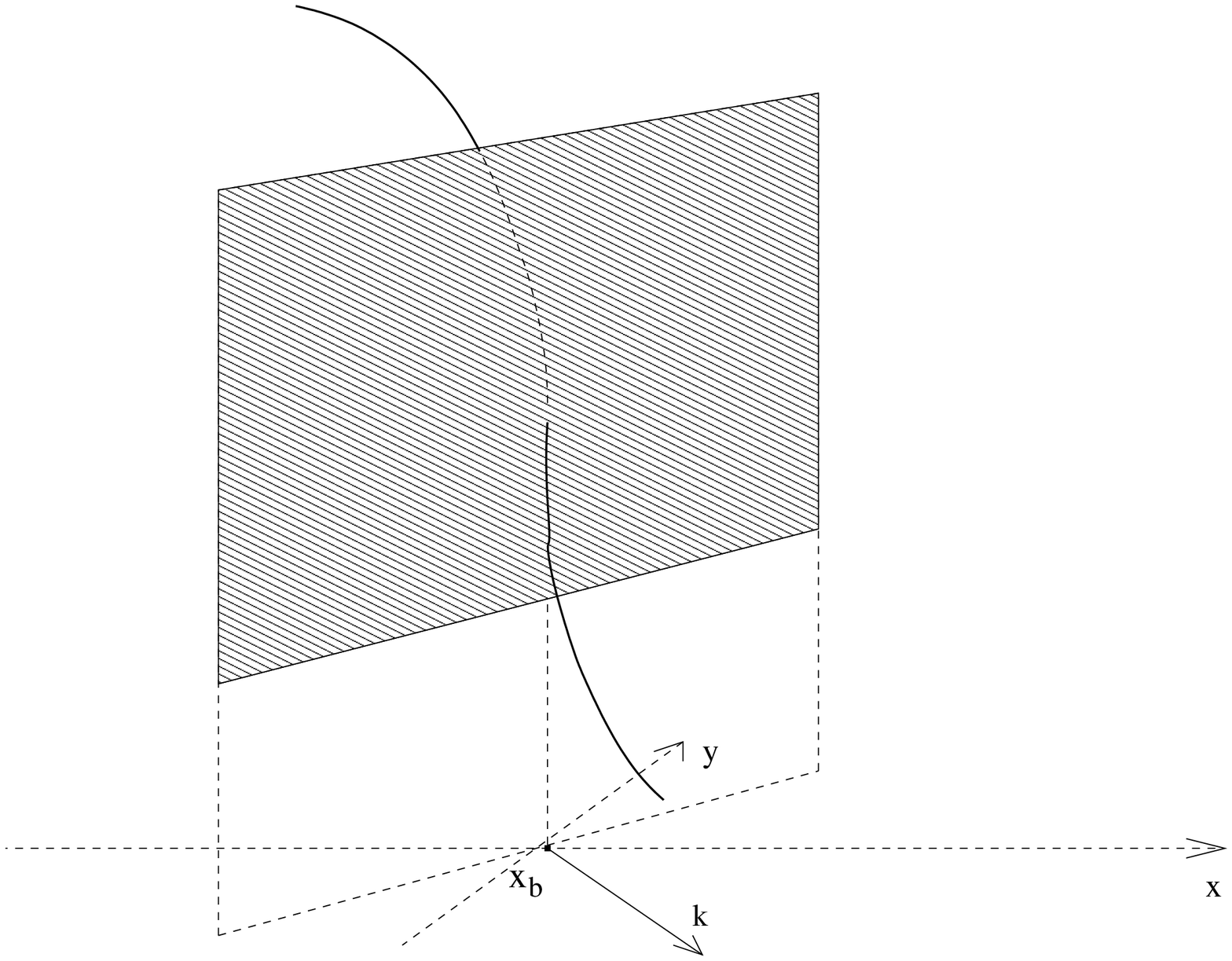}}
\end{center}
Fig.3 \;The vertical plane tangent to the wave at the breaking point.


\begin{center}
\mbox{ \epsfxsize=8cm \epsffile{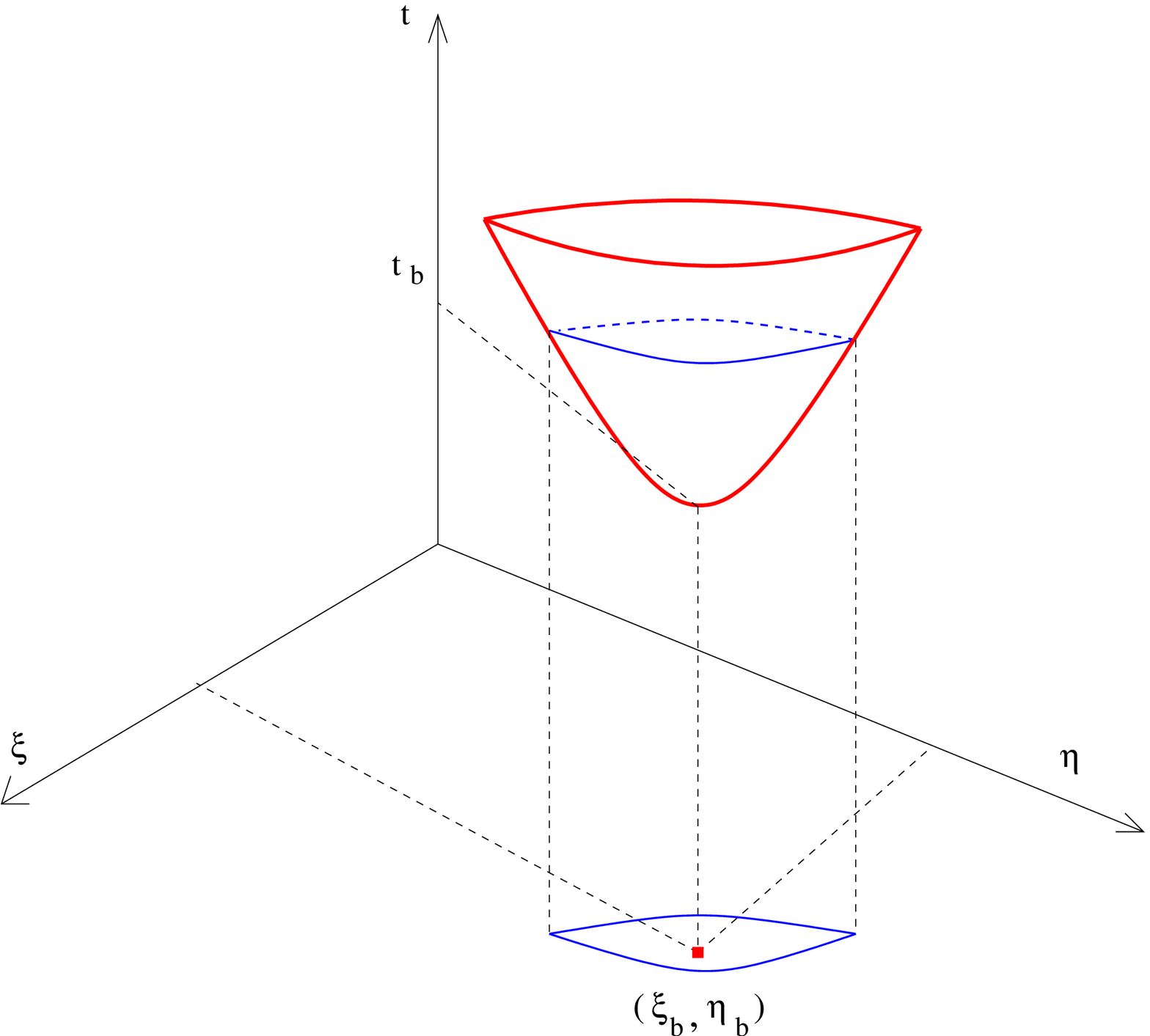}}
\end{center}
Fig.4. The two-dimensional singularity manifold ${\cal S}(\xi,\eta,t)=0$ near 
breaking.


\begin{center}
\mbox{ \epsfxsize=7cm \epsffile{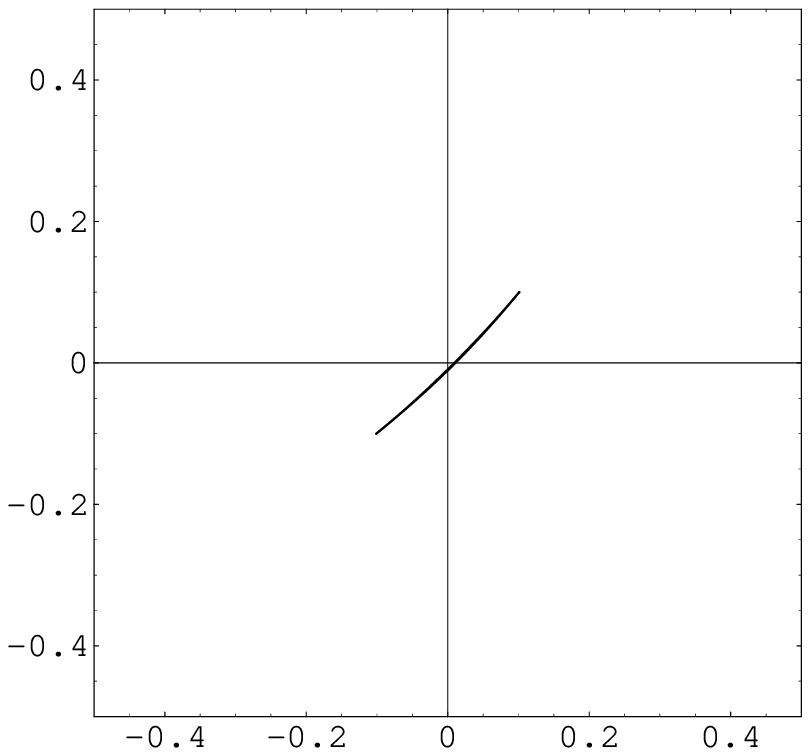}}\mbox{ \epsfxsize=7cm \epsffile{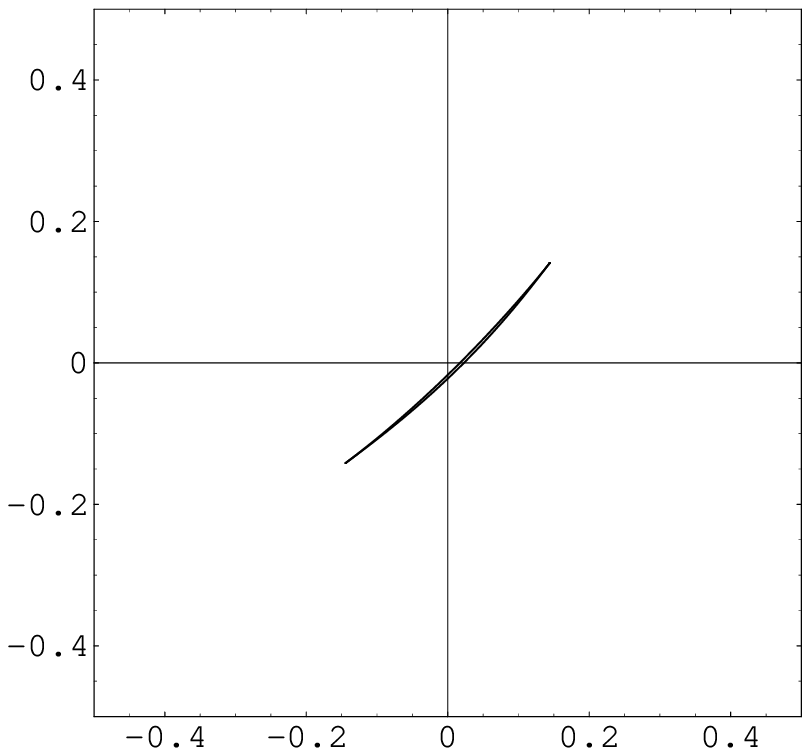}}
\mbox{ \epsfxsize=7cm \epsffile{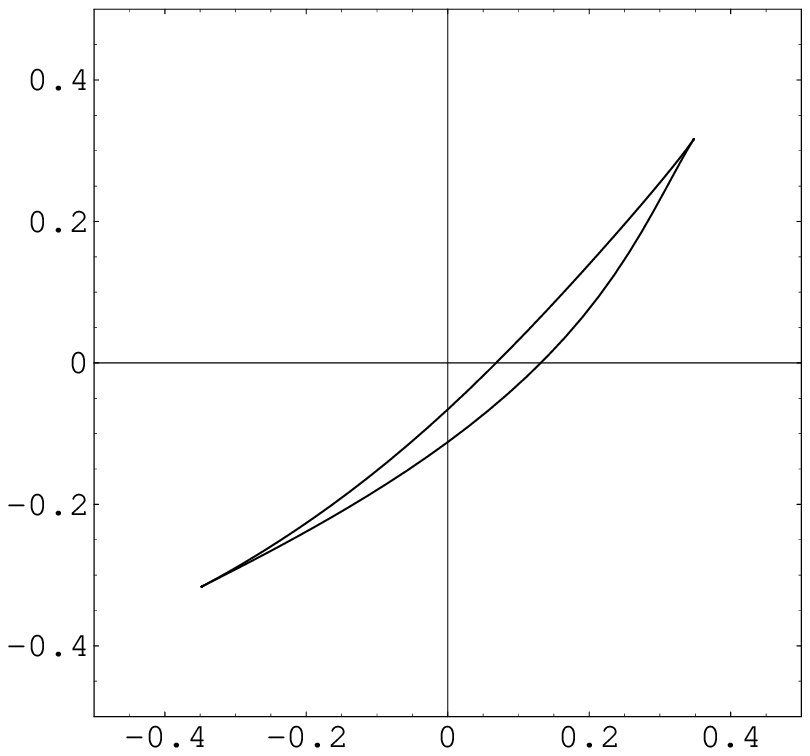}}\mbox{ \epsfxsize=7cm \epsffile{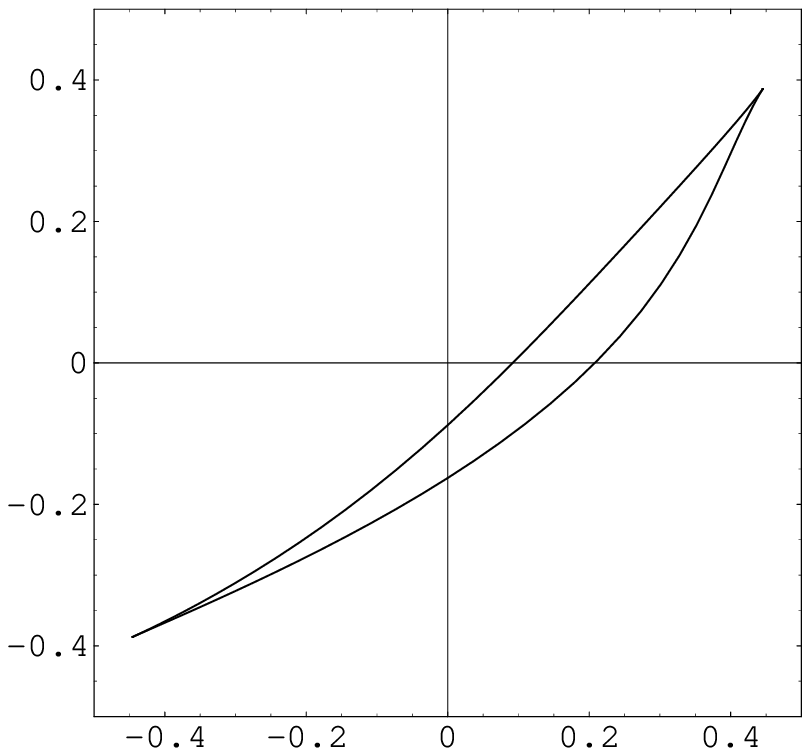}}
\end{center}
Fig.5. Four consecutive snapshots describing the time evolution of the three-valued region delimited 
by the caustic, immediately after breaking. The center of each picture is the breaking point. 


\begin{center}
\mbox{ \epsfxsize=8cm \epsffile{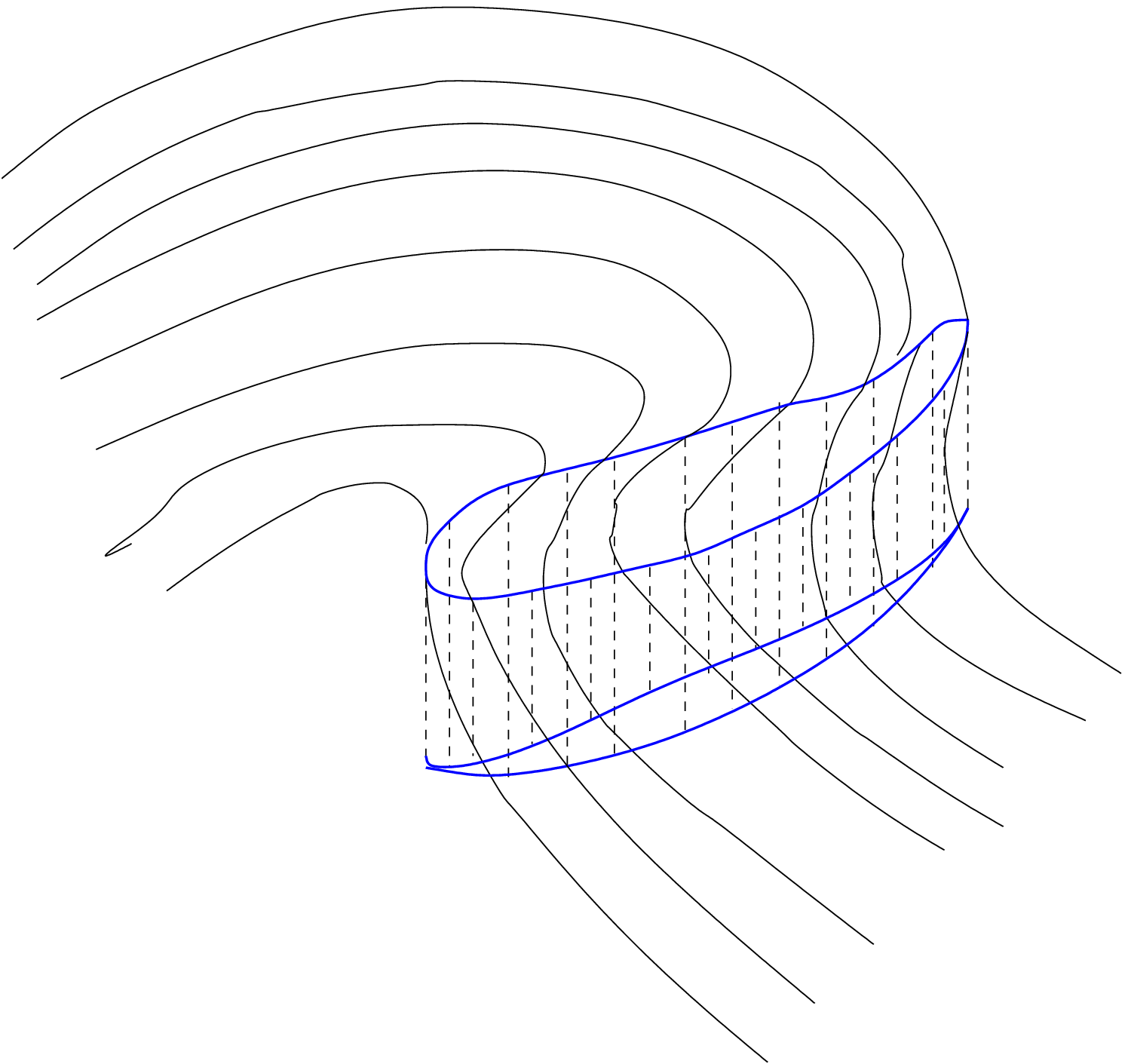}}
\end{center}
Fig.6. The three-valued solution after breaking, and the caustic delimiting the multivalued 
region.   


\begin{center}
\mbox{ \epsfxsize=9cm \epsffile{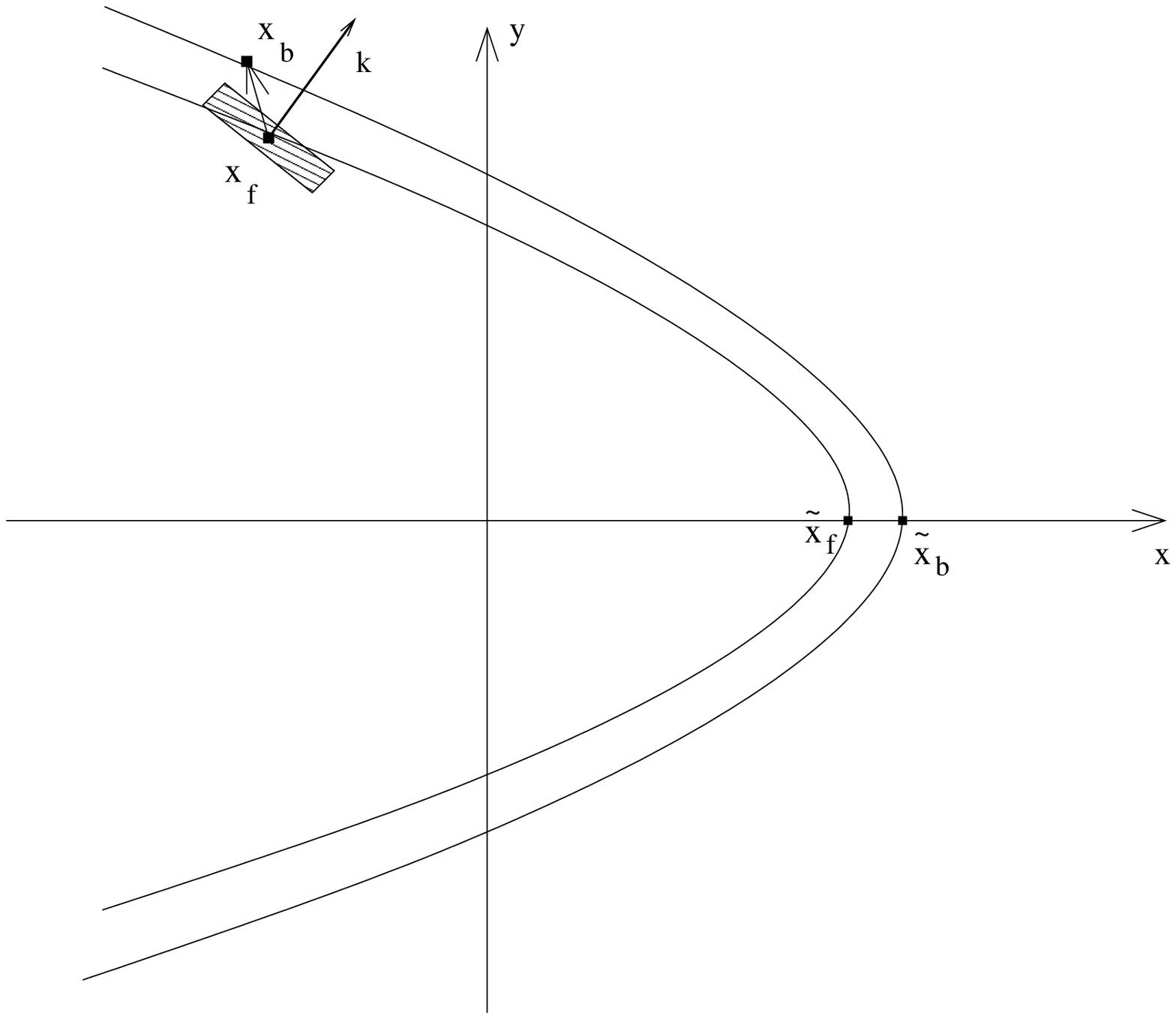}}
\end{center}
Fig.7. The two parabolas containing the inflection and breaking points, immediately 
before breaking, together with the narrow strip of the similarity regime.   

\vskip 10pt
\noindent
{\bf Acknowledgements}. This research has been supported by the RFBR 
grants 07-01-00446, 06-01-90840, and 06-01-92053, by the bilateral agreement 
between the Consortium Einstein and the RFBR, and by the bilateral agreement between the 
University of Roma ``La Sapienza'' and the Landau Institute for Theoretical Physics of the 
Russian Academy of Sciences. We thank B. Konopelchenko for pointing out some references 
related to our work, and E. A. Kuznetsov for useful discussions concerning the   
universality aspects of the gradient catastrophe.

\end{document}